\documentclass[amsmath,amssymb]{revtex4}

\usepackage{graphicx}
\usepackage{dcolumn}
\usepackage{bm}
\usepackage{multirow}
\usepackage{booktabs,siunitx}

\begin{document}

\preprint{APS/123-QED}%

\title{Mapping the magnetic state as a function of anti-site disorder \\ in Sm$ _{2} $NiMnO$ _{6} $ double perovskite thin films}

\author{Supriyo Majumder$^{a}$}
\author{Malvika Tripathi$^{a}$}
\author{R. Raghunathan$^{a}$}
\author{P. Rajput$^{b}$}
\author{S. N. Jha$^{b}$}
\author{D. O. de Souza$^{c}$}
\author{L. Olivi$^{c}$}
\author{S. Chowdhury$^{a}$}
\author{R. J. Choudhary$^{a}$}\email{ram@csr.res.in}
\author{and D. M. Phase$^{a}$}
\renewcommand{\andname}{\ignorespaces}
\affiliation{$^{a}$UGC DAE Consortium for Scientific Research, Indore 452001, India}
\affiliation{$^{b}$Beamline Development and Application Section, Bhabha Atomic Research Centre, Mumbai 400085, India}
\affiliation{$^{c}$Elettra Sicrotrone Trieste S.C.p.A., SS 14-km 163.5, 34149 Basovizza, Italy}


\begin{abstract}
The predictability of any characteristic functional aspect in a double perovskite system has always been compromised by its strong dependence over the inevitably present anti-site disorders (ASD). Here, we aim to precisely map the quantitative and qualitative nature of ASD with the corresponding modifications in observables describing the magnetic and electronic state in epitaxial Sm$ _{2} $NiMnO$ _{6} $ (SNMO) double perovskite thin films. The concentration and distribution patterns of ASD are effectively controlled by optimizing growth conditions and estimated on both local and global scales utilizing extended X-ray absorption fine structure and bulk magnetometry. Depending upon the defect densities, the nature of disorder distribution can vary from homogeneous to partially segregated patches. Primarily, the effect of varying B-site cationic arrangement in SNMO is reflected as the competition of long range ferromagnetic (FM) and short scale antiferromagnetic (AFM) interactions originated from ordered Ni-O-Mn and disordered Ni-O-Ni or Mn-O-Mn bonds, respectively, which leads to systematic shift in magnetic transition temperature and drastic drop in saturation magnetization. In addition, we have observed that the gradual increment in density of ASD leads to significant deviation from uniaxial anisotropy character, reduction in anisotropy energy and enhancement of moment pinning efficiency. However, the observed signatures of $ Ni^{2+}+Mn^{4+} \longrightarrow Ni^{3+}+Mn^{3+} $ charge disproportionation is found to be independent of cation disorder densities. This work serves as a basic route-map to tune the characteristic magnetic anisotropy, magnetic phase transitions, and magnetization reversal mechanism by controlling ASD in a general double perovskite system. 
\end{abstract}

\maketitle

\section{INTRODUCTION}
The periodic symmetry of any crystal structure is destined to be broken owing to imperfections and accordingly the physical properties of the system have to bear the consequences from slight to moderate level. Structural disorders have been well known to introduce new channels for charge transport \cite{RNoriega2013, AHusmann1996}, heat conductivity \cite{ABalandin2011}, magnetic exchange interactions \cite{JBowles2013} etc. The double perovskite A$_{2}$B$'$B$''$O$_{6}$ (where A is alkaline earth or rare earth cations and B$'$, B$''$ are transition metal cations in octahedral coordination with six oxygen anions) family is a relevant example of how the mis-locations of B$'$ / B$''$ ions from ideal alternating site occupancy, a phenomenon commonly known as anti-site disorder (ASD), can drastically alter the functional properties. In a perfectly ordered double perovskite configuration, the B$'$ and B$''$ cations are arranged in an ideal rock salt fashion which is kinetically favoured by significant differences in formal charge states and ionic radii \cite{MTAnderson1993}. The interactions among the ordered B-site cations are mediated via B$'$-O-B$''$-O-B$'$ long range chain, whereas, the presence of ASD results in two additional interaction pairs viz: B$'$-O-B$'$ and B$''$-O-B$''$, as illustrated in Fig.\ref{psd}(a). ASD plays a dominating role in stabilizing the electronic and magnetic ground states in the double perovskite systems and even a drastic transformation from ferromagnetic-metallic (FM-M) state to antiferromagnetic-insulating (AFM-I) state is possible in some specific cases due to presence of ASD \cite{MGHernandez2001, DDSarma2001}. However, how the concentration of ASD works together with their nature of distribution to govern properties of a general double perovskite system is still an open question. 

The class of Ni-Mn based R$_2$NiMnO$_6$ (RNMO, R: rare earth) double perovskite is specially prominent for realizing the exceptional FM-I state \cite{DChoudhury2012, NSRogadol2005, HJZhaonat2014, HJZhaoprb2014}. The rare combination of FM-I state in RNMO system is originated due to virtual hopping of the electrons from half filled e$ _{g} $ orbital of Ni to empty e$ _{g} $ orbitals of Mn leading to Ni(e$ ^{2}_{g} $)-O-Mn(e$ ^{0}_{g} $) super exchange interactions, which is FM in 180$ ^{o} $ geometry according to GK rule \cite{GoodenoughKanamoril195559}. FMIs can save the energy loss due to large eddy currents and charge displacement and hence, essential for modern day dissipationless quantum electronics, spin-wave based spintronics and information processing in solid state quantum computing \cite{CSohn2019, JFrantt2019, YFujioka2019}. In addition, the recent observations of near room temperature spin pumping, giant magneto-dielectricity, magneto-capacitance, magneto-resistance, multiferroic and magneto-electric effects has rejuvenated interest in the fundamental aspects of double perovskites with Ni and Mn as B-site cations \cite{YShiomi2014, DChoudhury2012, NSRogadol2005, MAzuma2005, JSu2015}. However, the small difference in ionic radii of Ni and Mn cations makes the presence of ASD practically unavoidable in RNMO system. That is why the reproducibility and reliability of several physical properties of RNMO family has been questionable and as a consequence, the utilization of RNMO double perovskite in device fabrication is still a challenging issue. 

One of the key reasons for these discrepancies is the lack of a general functional relationship that maps the distinct features of ASD to the corresponding modification in microscopic configuration resulting in a particular electronic and magnetic ground state. An experimental approach to develop such a relationship will involve following steps: to find a method for fine tuning the disorder fraction; precise quantification of ASD concentration and the nature of disorder distribution; and to observe the exclusive modification in ground state of the system due to ASD. The accurate estimation of ASD density is crucial to understand the disorder mediated effects and to engineer the multifunctional properties of double perovskite systems for technological feasibility. Although the presence of ASD is intrinsic, to precisely estimate the fraction and predict the qualitative nature of these disorders in a complex crystal environment is not a straightforward task. Fine tuning the disorder in a controlled fashion can be achieved by fabricating double perovskite systems in thin film form, which provides a wide range of control parameters during synthesis. With this motive, we have synthesized epitaxial thin films of Sm$_{2}$NiMnO$_{6}$ (SNMO) to generate different ASD densities by varying deposition parameters for investigating the role of B-site cation disorder on electronic and magnetic states. Utilizing microscopic (extended X-ray absorption fine structure) and macroscopic (bulk magnetometry) probes duly combined with structure simulations (random alloy) we are able to determine the degree and distriribution nature of cation disorder present in the system. Among the trivalent rare-earth elements, Sm based systems show unique trends in magnetization and anisotropy behaviors attributed to narrow energy separation ($ \sim $0.12 eV) between the ground state (J=5/2) and lower excited states (J=7/2, 9/2 etc.) of Sm$ ^{3+} $ ions leading to significant mixing of ground state with excited multiplets in presence of crystal field, exchange field and/or thermal energy \cite{KHJBuschow1974}. The presence of Sm ion is expected to introduce additional temperature dependent perturbations in the long-range ordered Ni-Mn networks, making the ground state more complex and challenging to understand. In this work, here we demonstrate a pathway to tailor the magnetic ground state, phase transition temperature, magnetic anisotropy character and moment pinning efficiency by engineering the octahedral site cation arrangement in the system.

\section{EXPERIMENTAL METHODOLOGY}
SNMO thin films of thickness $\sim$140$\pm$5 nm were fabricated on (001) oriented single crystalline SrTiO$ _{3} $ (STO) substrates using pulsed laser deposition system equipped with KrF excimer laser (Lamda Physik, wavelength $ \lambda $=248 nm, pulse width 20 ns), external focusing optics, vacuum chamber and \textit{in situ} reflection high energy electron diffraction (RHEED) assambly (Staib Instruments, kSA400). For the ablation process, stoichiometric polycrystalline bulk target of SNMO was synthesized by conventional solid state reaction method. During the film growth, the phase stability was monitored while varying the deposition temperature (DT) within the range of 650-780 $ ^{o} $C and oxygen partial pressure (OPP) within the range of 200-800 mTorr. The laser fluence at the target surface was set to 2 J-cm$ ^{-2} $ and target to substrate distance was kept constant at 4.5 cm during all the depositions. Just after deposition, initially films were annealed for 5 minutes at same temperature as used during growth process followed by cooling under 400 Torr OPP with different cooling rates ranging between 10-40 $ ^{o} $C/min. The structural phase of the thin films was characterized by X-ray diffraction $2\theta$-scans (Bruker D2 Phaser Desktop Diffractometer), X-ray rocking $ \omega $-scan and Reciprocal space mapping (RSM) studies (High Resolution X-ray Diffractometer Bruker D8 Discover) using Cu K$ \alpha $ ($ \lambda$ = 1.54 $ \AA $) source. Chemical valence state of the elements present in the sample were investigated by X-ray photo electron spectroscopy (XPS) experiments using Al \textit{K}$ _{\alpha} $ ($ h\nu $ = 1486.7 eV) lab-source and hemispherical energy analyzer (Omicron, EA-125, Germany) at Angle Integrated Photoemission Spectroscopy (AIPES) beamline (Indus-1, BL 2, RRCAT, Indore, India) and X-ray absorption near edge spectroscopy (XANES) measurements performed in fluorescence mode using hard X-ray synchrotron radiation (Indus-2, BL 9, RRCAT, Indore, India). During XPS measurements experimental chamber vacuum was of the order of $ 10^{-10} $ Torr. The charging effect corrections in XPS were accounted by measuring C 1\textit{s} core level spectra. XPS spectra were de-convoluted by fitting with combined Lorentzian-Gaussian function and Shirley background using XPSPEAK 4.1 program. The estimated energy resolution ($ \Delta E / E $) for XPS and XANES measurements across the measured energy range were about 6$ \times $10$ ^{-4} $ and 1$ \times $10$ ^{-4} $, respectively. Local coordination environment of the thin film samples was probed using extended X-ray absorption fine structure (EXAFS) measurements at Ni \textit{K}-edge. EXAFS spectra of thin films were recorded in fluorescence mode at the XAFS beamline (11.1R, Elettra-Sincrotrone, Trieste, Italy). The presence of Sm \textit{L}$_{3}$-edge (6716 eV) in the vicinity of Mn \textit{K}-edge (6539 eV) restricts to perform EXAFS measurements at Mn \textit{K}-edge. Reference absorption edge spectra of metal foils (Mn and Ni) were used for energy calibration of the incident X-ray in XANES and EXAFS measurements. In order to get better statistics, the EXAFS scans for each sample were collected five times and merged. Then standard background subtraction and normalization procedures were applied to extract normalized XANES spectra and EXAFS oscillations in $k$-space ($ \chi(k) $) using ATHENA program \cite{BRavel2005} and implementing AUTOBK algorithm \cite{MNewville1993}. Fourier transformations of the EXAFS signal (within selected $k$ range) were calculated to observe the R-space spectra($ \chi(R) $). Thereafter, obtained EXAFS spectra were fitted with specific model implementing ARTEMIS software which uses ATOMS and FEFF6 programs \cite{BRavel2001, JJRehr200009} to simulate theoretical spectrum by summing over all partial contribution from scattering paths for a given crystallographic structure (Table. S1 in Supplementary Material (SM) \cite{SM}). Each contribution (for example, consider $ i $th coordination shell), was computed using standard EXAFS expression \cite{JJRehr200009}, with refinable structural parameters: coordination number ($ N_{i} $), average coordination distance ($ R_{i} $) and mean-square relative displacement (MSRD) factor ($ \sigma_{i}^{2} $). All of the above described spectroscopic measurements were carried out at 300 K. Magnetization measurements were performed using MPMS 7-Tesla SQUID-VSM (Quantum Design Inc., USA). Temperature dependent magnetization M(T) was measured following conventional zero field cooled (ZFC) warming and field cooled warming (FCW) protocols. Before all the M(T) measurements the samples were heated above their respective magnetic ordering temperatures in order to remove prior history and standard de-Gaussing procedure was followed to remove the trapped magnetic field inside the superconducting magnet of the magnetometer. Magnetic moment calibration was checked by measuring data for standard palladium sample. In our measurements, the background moment contribution was of the order of $ \sim 10^{-8}$ emu as shown in SM \cite{SM} Fig. S3(a). To be more precise about the magnetic moment contribution from the thin films, the temperature dependent magnetization in presence of applied magnetic field $ \mu_{0} $H=100 Oe is measured for the empty sample holder assembly, bare STO (001) substrate and STO$ \mid $SNMO samples mounted on sample holder (as presented in SM \cite{SM} Fig. S3(a, b)). Observed moment values at T=10 K in presence of $ \mu_{0} $H=100 Oe, are listed in SM \cite{SM} Table. S3. It was quite evident from these behaviors that the background and substrate contribution was significantly low as compared to the film contribution, confirming the observed magnetic results for the samples are intrinsic magnetic properties of SNMO system. However, as diamagnetic moment is linear in field (with a negative sign), at very high magnetic field values where sample ferromagnetic moments are saturated (constant), the substrate moment signal continues to increase with increasing magnetic field. The diamagnetic contribution from the substrate was calculated from the negative slope of the linear part at high field values of the isothermal magnetization versus field M(H) curves, as displayed in SM \cite{SM} Fig. S3(c) and subtracted from the overall magnetic moment of the samples.  

\section{RESULTS AND DISCUSSION}
\begin{figure*}[t]
\centering
\includegraphics[angle=0,width=1.0\textwidth]{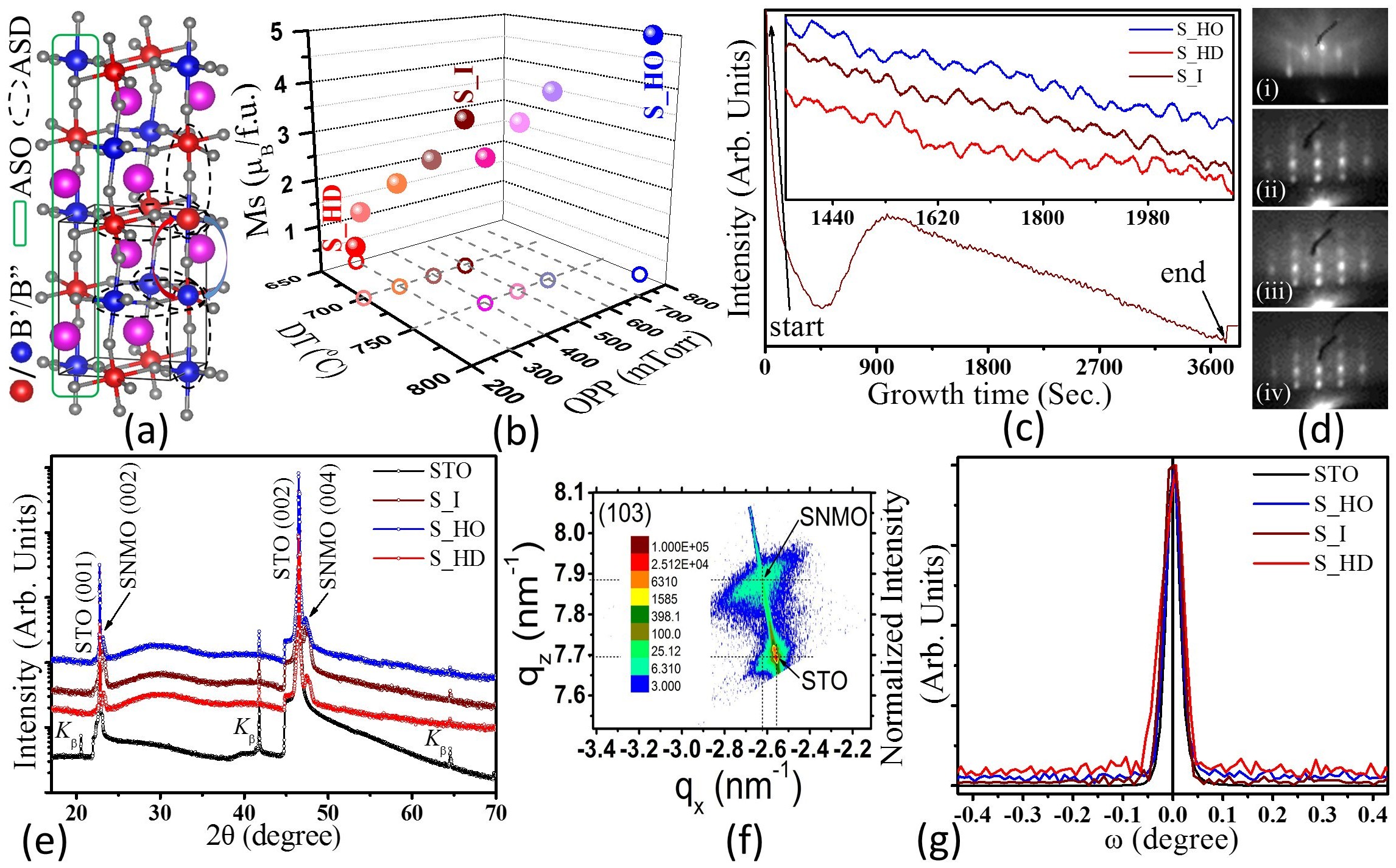}
\caption{(a): Schematic visualization of A$ _{2} $B$'$B$''$O$ _{6} $ (large magenta spheres: A-site cations, Red/Blue sphere: B$'$/B$''$ cations and small grey spheres: oxygen anions) double perovskite crystal structure demonstrating; anti-site order (ASO): ideal alternating arrangement of B$'$/B$''$ cations (highlighted by solid/green box) leading to B$'$-O-B$''$ long range ordered chains and anti-site disorder (ASD): mislocation (as shown by arrow) of B-site cations which results in disordered B$'$-O-B$'$, B$''$-O-B$''$ pairs (highlighted by dash/black oval). (b): Phase stability mapping for the growth of SNMO films, illustrating the tunability over ASD phase fraction via controlling growth parameters. Hollow circles correspond to projection of same colored solid spheres in deposition temperature - oxygen partial pressure plane. (c): Representative \textit{in situ} reflection high energy electron diffraction (RHEED) specular beam intensity evolution in full growth time window for S$ \_ $I thin film deposition. Inset compares RHEED oscillations for S$ \_ $HO, S$ \_ $I and S$ \_ $HD film growths. (d): RHEED patterns along (001) direction for (i) substrate, (ii) S$ \_ $HD, (iii): S$ \_ $I and (iv): S$ \_ $HO thin film surfaces. (e): X-ray diffraction $2\theta $ patterns of SNMO thin films deposited under different conditions on (00l) STO substrate along with diffraction profile for the (00l) STO substrate. Observed diffraction peaks from the films at just right side of the substrate peaks are marked by arrows. Peaks at the left side of the substrate peaks corresponds to (00l) K$ _{\beta} $ reflections. (f): Representative reciprocal space map around asymmetric (103) plane for S$ \_ $I film. The most intense coordinates in (q$ _{x} $, q$ _{z} $) domain correspond to SNMO film and STO substrate are identified by arrows. (g): X-ray rocking $ \omega $-scans across (00l) crystallographic plane of STO substrate and SNMO films.}\label{psd}
\end{figure*}

\subsection{Thin film fabrication}
The growth conditions during the thin film deposition process, especially the values of DT and OPP, significantly influence the cation ordering in a double perovskite structure. One of the key signature for degree of B-site cationic ordering is the saturation magnetization \cite{RIDass2003, DYang2018, DYang2019}. We have utilized the value of saturation magnetization $M_{S}$(T=5 K) measured as a function of DT and OPP to map the cation-ordering phase diagram for SNMO thin films, as presented in Fig.1(b). The phase mapping depicts that although SNMO can be stabilized in single phase (discussed in Subsection: Global structure) for a broad growth parameter window, but the films fabricated under low DT (650$\pm$30 $^{\circ}$C) and low OPP (300$\pm$100 mTorr) conditions have highly disordered phase. With increasing DT and OPP the level of cation ordering improves and finally highly ordered phase can be achieved with high DT (780$\pm$30 $^{\circ}$C) and high OPP (800$\pm$100 mTorr). Under a fixed OPP value, the rise in DT enhances the B-site ordering in the lattice and vice-versa, for a fixed DT value the increase in OPP value also helps in stabilizing the more ordered phase. The value of DT used during fabrication process plays an important role for the film growth kinetics and atomic arrangement \cite{XYang2017, SEKaczmarek2017}. Higher DT may allow the cations for ordered occupancy in the B-site sub-lattice via thermal diffusion. On the other hand, varying the value of OPP affects the flux and energy of the impinging species during film growth process \cite{HZGuok2008, JPMaria1998}. For the higher OPP values, there may be a reduction in the defects associated with the bombardment of high energy particulates. Thus, we have found that different levels of anti-site cation ordering can be engineered by fine tuning of film growth parameters. To investigate the ASD mediated modulation in electronic and magnetic properties of SNMO system, we have fabricated three films namely: (i) S$\_$HO with highly ordered phase (DT = 780 $^{\circ}$C, OPP = 800 mTorr), (ii) S$\_$HD having highly disordered phase (DT = 650 $^{\circ}$C, OPP = 300 mTorr); and (iii) S$\_$I with an intermediate phase which have admixture of both ordered and disordered structures (DT = 700 $^{\circ}$C, OPP = 500 mTorr). 

The thin film growth dynamics is monitored by continuous recording \textit{in situ} real time RHEED patterns. The evolution of RHEED specular beam intensity as a function of growth time is displayed in Fig. \ref{psd}(c). Within entire elapsed time window, RHEED oscillations persevere uniform amplitude superimposed on a damping background which suggests island (3D) type (Volmer-Weber) of film growth mode. Figure \ref{psd}(d) compares RHEED patterns for (i) thermally cleaned and oxygen radical pretreated substrate surface with (ii-iv) thin film surfaces at the end of growth process. The sharp streaky RHEED pattern from substrate indicates atomically flat 2D surface, whereas observed spots for the grown films point out 3D island type of surface. Thin film RHEED patterns recorded with in-plane rotations as displayed in SM \cite{SM}, indicate single domain epitaxial growth of SNMO system. It is worth to mention here that SNMO crystal adopts same growth mode for S$ \_ $HO, S$ \_ $I and S$ \_ $HD thin film depositions as evidenced in Inset of Fig. \ref{psd}(c). This is because of same misfit strain (discussed later) from the substrate which plays the dominating role in stabilizing the grown overlayer film. 

Under pseudo-cubic approximation, SNMO lattice can be described as $ \sqrt{2}a_{p} \times \sqrt{2}a_{p} \times 2a_{p} $, where $a_{p}$ represents pseudo-cubic lattice parameter. The average lattice parameter of bulk SNMO is $ a^{bulk}_{p} $ = 3.8373 \AA, which is smaller than STO (cubic \textit{Pm$\bar{3}$m} symmetry) lattice parameter $ a^{sub}_{p} $ = 3.905 \AA. Calculated lattice mismatch with the substrate defined as ($a_{bulk}$-a$_{sub}$)/a$_{bulk}$ is $ \sim $-1.76\%. During the initial growth stage the lattice misfit between grown epitaxial thin film and under-layer substrate is compensated by incorporating misfit strain. Associated strain energy accommodated in strained film layer increases in proportion with increasing thickness of the over-layer film. When the thickness reaches critical value to sustain strain state, misfit defect is formed to partially release stored strain energy. The substrate induced strain can be relaxed through 3D island formation as it is easier to expand or contract the lattice for islands than flat layers \cite{AFluri2018}. 

\subsection{Global structure}
The X-ray diffractograms of SNMO thin films on (00l) oriented single crystal STO substrates are presented in Fig. \ref{psd}(e). Independent of growth parameter variation (within the range used here), samples show diffraction peaks corresponding to SNMO (monoclinic \textit{P2$  _{1}$/n} symmetry) phase (00l) reflections only, i.e. the grown films are in single phase and (00l) oriented. We have not observed any discernible change in XRD 2$ \theta $ patterns for the films having different level of cation disorder, confirming similar SNMO average crystal structure in all the samples. This is expected as Ni and Mn transition metals have comparable ionic radius and therefore any local interchange between Ni and Mn cations from their respective site occupancy does not affect the average crystal structure of the SNMO system. However, the different level of cation occupancy defects present in SNMO films will certainly leave their imprints on local structure (as discussed in Section: Local structure). The out-of-plane lattice parameter $ c $ calculated from (00l) reflections, are $ \sim $7.608($ \pm0.003 $) $ \AA $ for S$ \_ $HO, $ \sim $7.609($ \pm0.003 $) $ \AA $ for S$ \_ $I and $ \sim $7.605($ \pm0.003 $) $ \AA $ for S$ \_ $HD. SNMO bulk cell constant $ c $ (Table S1 in SM \cite{SM}) matches well with these film out-of-plane lattice parameter values. 

To explore the possibility of substrate induced strain present in the films, Reciprocal space maps across asymmetric (103) reflection are recorded as shown in Fig. \ref{psd}(f). RSM measurements confirm the epitaxial nature of SNMO films. The q$_{x} $ and q$_{z} $ axis shown in RSM plot correspond to the in-plane and out-of-plane directions, respectively. Here, the maximum intensity coordinate for the film (q$_{xf} $, q$_{zf} $) Bragg's spot is shifted from the substrate (q$_{xs} $, q$_{zs} $) case, indicates that the films are relaxed from substrate clamping effects. Consequently, the lattice parameters of SNMO film along in-plane and out-of-plane directions, estimated from q$_{xf} $ and q$_{zf}$, are found close to the SNMO bulk value. The strain relaxation may locally break long range periodicity of the lattice by introducing misfit defects via several forms, for example: edge dislocation, oxygen vacancy channel etc. 

The X-ray rocking $ \omega $-scan is a widely used technique to study distribution of dislocation defects in single crystals. Dislocations cause broadening in rocking curve and dislocation density ($ \rho $) is approximated as \cite{PGay1953}, $ \rho = {(\Delta \omega)^{2} / (9b^{2})} $ considering their random distribution. Here, $ \Delta \omega $ is angular spread of X-ray diffraction from a particular (hkl) plane and $ \textbf{b} $ is Burgers vector defined as lattice distortion caused by dislocation defect. For instance, in cubic lattice, the magnitude of $ \textbf{b} $ resulting from edge dislocation, is similar to interatomic distance of that lattice. Therefore, the full width at half maximum (FWHM) of rocking curve is directly related to dislocation defect distribution in grown film. Figure \ref{psd}(g) presents rocking $ \omega $ scans for epitaxial SNMO film (004) reflection along with STO single crystal substrate (002) reflection. Estimated FWHM values for the films, S$ \_ $HO: 0.04($\pm$0.01)$ ^{o} $, S$ \_ $I: 0.04($\pm$0.01)$ ^{o} $, S$ \_ $HD: 0.05($\pm$0.01)$ ^{o} $, are very close to substrate, STO: 0.03($\pm$0.01)$ ^{o} $ value, suggesting good crystallinity of the grown films, similar to single crystal substrate. Negligibly small relative differences in FWHM of S$ \_ $HO and S$ \_ $HD samples, which is comparable to instrumental resolution limit, indicate no change of dislocation defect distribution in different SNMO films.

\begin{figure*}[t]
\centering
\includegraphics[angle=0,width=0.8\textwidth]{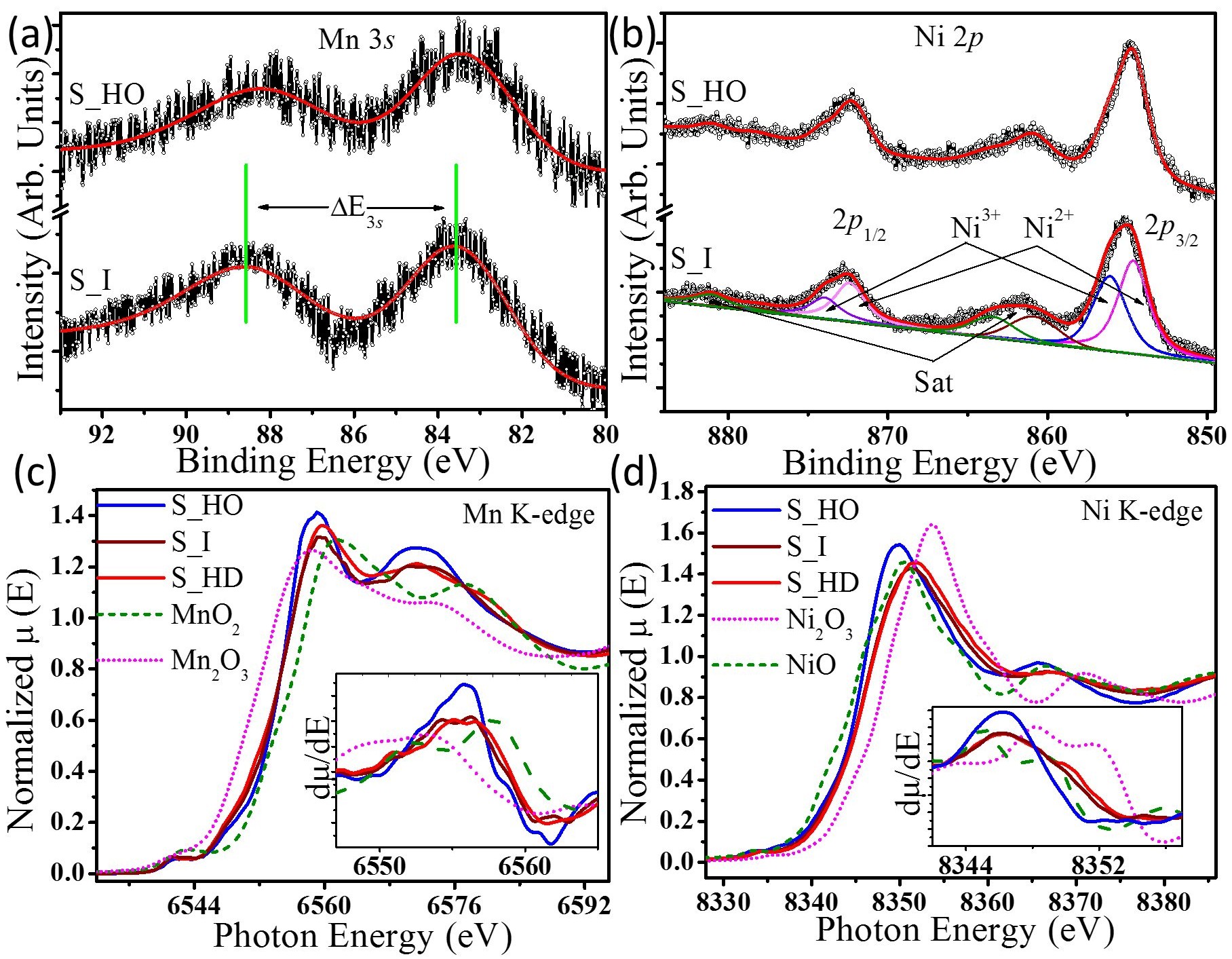}
\caption{X-ray photoemission: (a) Mn 3\textit{s}, (b): Ni 2\textit{p} core level spectra and X-ray absorption: (c) Mn, (d) Ni \textit{K}-edge spectra measured for SNMO films, suggesting mixed valency of both Mn (4+/3+) and Ni (2+/3+) species. Insets of (c, d) present first order derivative of near edge X-ray absorption spectra.}\label{XPS XANES}
\end{figure*}

\subsection{Core-level electronic structure}\label{Sec Core-level electronic structure}
We have recorded photo-emission spectra utilizing soft X-rays and near-edge absorption spectra with hard X-rays in order to probe the charge states of constituting elements present at the surface as well as in the bulk portion of the thin films.  

The multiplet energy splitting in the 3\textit{s} core XPS spectra is used to determine the valance state of Mn species at surface. After emission of one photo electron from 3\textit{s} core level, Mn ions have two multiplets originated due to exchange coupling between remaining core electron at 3\textit{s} level and the electrons at unfilled 3\textit{d} shell \cite{NMannella2008}. The magnitude of spectral splitting is estimated as
\begin{equation}\label{EqDE3s}
\Delta E_{3\textit{s}} = (2s+1) J^{eff}_{3\textit{s}, 3\textit{d}}
\end{equation}
where s is net valence spin of emitter specie, J$ ^{eff}_{3\textit{s}, 3\textit{d}} $ is effective exchange integral between 3\textit{s} and 3\textit{d} states considering final state intra-shell correlations \cite{NMannella2008, LSangaletti1995}. Equation \ref{EqDE3s} provides a method to estimate the total valence spin of emitter ions using the 3\textit{s} spectral splitting. In case of Mn, the 3\textit{s}-3\textit{d} exchange interaction is the most dominating effect to govern the splitting in 3\textit{s} spectra \cite{NMannella2008, LSangaletti1995}. Therefore, Mn 3\textit{s} multiplet splitting can be used as a sensitive probe to determine the valency of Mn species present in the sample. However, this method does not help in accurate determination of valence state for heavier transition metals with high \textit{d} electron count (e.g. Ni) and also for lower electronegative ligands (e.g. Br), in which charge transfer final state screening effect increases \cite{NMannella2008}. The difference between the Mn 3\textit{s} XPS spectra for all the SNMO films (Fig.\ref{XPS XANES}(a)) is under experimental resolution. The observed multiplet splitting energy $ \Delta $E$ _{3\textit{s}} \sim $ 4.9 eV is in between $ \Delta $E$ _{3\textit{s}} $(Mn$ ^{4+} $) $ \simeq $ 4.4 eV and $ \Delta $E$ _{3\textit{s}} $(Mn$ ^{3+} $) $ \simeq $ 5.5 eV which suggests that Mn species have mixed valence character in SNMO films. For mixed valence Mn 3\textit{d}$ ^{(4-x)} $ configurations [where x Mn$ ^{4+} $ + (1-x) Mn$ ^{3+} $ = Mn$ ^{(3+x)+} $], the net valence spin is s = (1/2)(4-x). Therefore, from Eq.\ref{EqDE3s} the multiplet energy separation will be $ \Delta $E$ _{3\textit{s}} \simeq $ 1.1(5-x) where J$ ^{eff}_{3\textit{s}, 3\textit{d}} \simeq $ 1.1 eV as observed in previous studies on Mn based compounds \cite{LSangaletti1995, VRGalakhov2002}. From this obtained mixed valence fractional concentration in SNMO samples (having $ \Delta $E$ _{3\textit{s}} \sim $ 4.9 eV) are as follows, 54($ \pm $2)\% Mn$ ^{4+} $, 46($ \pm $2)\% Mn$ ^{3+} $.

Ni 2\textit{p} core level spectra splited into 2\textit{p}$ _{1/2} $ and 2\textit{p}$ _{3/2} $ due to spin-orbit splitting are shown in Fig.\ref{XPS XANES}(b). The nominal shift observed in these spectra for different SNMO samples can be ignored as they are under resolution limit. The asymmetry and broadening of spectral characters suggest the presence of more than one valence features in Ni species. Taking into account this asymmetry and broadening behavior, the Ni 2\textit{p} XPS spectra is deconvoluted with peaks centered at 854.6 eV (Ni$^{2+}$ 2\textit{p}$ _{3/2} $), 856.1 eV (Ni$^{3+}$ 2\textit{p}$ _{3/2} $) \cite{APGrosvenor2006}; 872.2 eV (Ni$^{2+}$ 2\textit{p}$ _{1/2} $) and 874.2 eV (Ni$^{3+}$ 2\textit{p}$ _{1/2} $). Mixed valence concentration evaluated from integrated intensity of fitted peaks reavels contribution from 55($ \pm $1)\% Ni$ ^{2+} $, 45($ \pm $1)\% Ni$ ^{3+} $. According to cluster model approximation the ground state of late transition metal central cation surrounded by ligand anions is described by \cite{AEBocquet199295}, 
\begin{equation}\label{EqCMg}
\Psi_{g} = a_{0} \vert \textit{d}^{n} \rangle + \sum_{m} a_{m} \vert \textit{d}^{n+m} \underline{L}^{m} \rangle : m = 1,..10-n
\end{equation}
where $ \underline{L} $ represents a hole in ligand band. After 2\textit{p} core level photo emission this will have final state consisting of screened state $ \underline{c} 3 \textit{d}^{n+1}\underline{L} $ and satellite states with $ \underline{c} 3 \textit{d}^{n} $ or $ \underline{c} 3 \textit{d}^{n+2}\underline{L}^{2} $, where $ \underline{c} $ refers to a core hole \cite{AEBocquet199295}. This approach explains the presence of satellite features (Fig.\ref{XPS XANES}(b)) in Ni$ ^{2+/3+} $ (with 3\textit{d}$ ^{8/7} $ configurations) 2\textit{p} core level spectra.

To further investigate whether these mixed valence nature are only surface properties or similarly distributed in bulk portion of the sample as well, we have used the large penetration depth of hard X-rays and measured the XANES spectra across Mn and Ni \textit{K}-edges (Fig.\ref{XPS XANES}(c, d)) for SNMO films along with standard samples of MnO$ _{2} $(Mn$ ^{4+} $), Mn$ _{2} $O$ _{3} $(Mn$ ^{3+} $), NiO(Ni$ ^{2+} $), Ni$ _{2} $O$ _{3} $(Ni$ ^{3+} $). XANES spectra at transition metal \textit{K}-edge is ascribe to photo electron excitation from atomic 1\textit{s} to empty \textit{p}-bands. The pre-edge features can be attributed to 1\textit{s}-3\textit{d} quadrapole or weakly allowed dipole transitions which arise possibly due to octahedral distortion related \textit{p-d} hybridized states \cite{FBridges2001}. Both of the XANES spectra recorded at Mn and Ni \textit{K} edges show minor pre-edge signatures, the characteristic white line just near the absorption edge and post edge oscillation. It is known that, with increasing valency of transition metal ions the band edge shifts to higher energies \cite{AHdeVries2003}. The absorption edge energy, defined as the first maxima in energy derivative of absorption spectra or the first inflection point in $ \mu $(E) are almost same in the three SNMO samples. Comparing the SNMO transition metal (Mn/Ni) \textit{K}-edge XANES spectra with the corresponding \textit{K}-edge spectra of standard samples (insets of Fig.\ref{XPS XANES}(c, d)), it can be inferred that Mn and Ni have valency in between 3+, 4+ and 2+, 3+ respectively. This observation is well consistent with our previously discussed XPS results. 

\begin{figure*}[t]
\centering
\includegraphics[angle=0,width=0.98\textwidth]{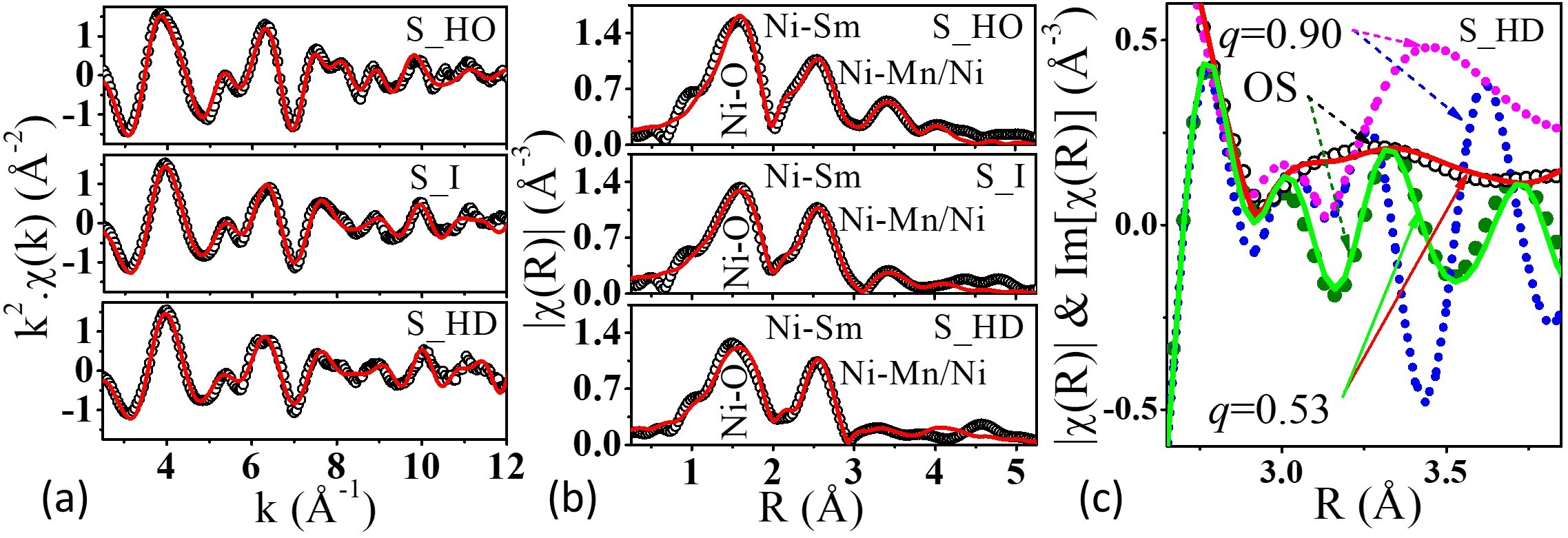}
\caption{Ni \textit{K}-edge extended X-ray absorption fine structure analysis, (a): $ k^{2} $-weighted spectra ($ k^{2} . \chi(k) $) and corresponding (b): modulus ($ | \chi (R) |$) of the Fourier transforms presenting observed signals (hollow/black circles) and their best fits (solid/red lines) for the SNMO thin films. Contributions from different coordination shells are identified in different regions in the spectra. (c): For S$ \_ $HD sample $ | \chi (R) |$ (hollow/black circles: observed data, solid/red line: best fit, dots/magenta: $ q $ = 0.9 spectra) and imaginary part ($ Im [\chi (R)] $) of the Fourier transforms (solid/dark green circles: observed data, solid/light green line: best fit, dots/blue: $ q $ = 0.9 spectra) show comparison of the OS: observed signal with the best fitted curve obtained by refining $ q $ (= 0.53) and a curve with forcefully fixed wrong $ q $ (= 0.9).} \label{EXAFS}
\end{figure*}

Therefore, mixed valence nature for both Ni$ ^{2+/3+} $ and Mn$ ^{4+/3+} $ species are confirmed from photo emission as well as photo absorption measurements. On the other hand, Sm is found in 3+ valence state from both XPS and XANES studies (not shown here). O \textit{K}-edge probes the unoccupied density of states just above the Fermi level involving O 2\textit{p} - transition metal 3\textit{d} hybridized bands. Due to hybridization, the local crystal field from surrounding lattice O ligands, splits the transition metal \textit{d} bands into t$ _{2g} $ and e$ _{g} $ states \cite{deGroot1989}. Any O vacancy in the lattice introduces a change in crystal field symmetry of O ions as well as electron population of transition metal \textit{d} states. As a result, the relative intensity of t$ _{2g} $ and e$ _{g} $ features in O \textit{K}-edge XANES is highly sensitive to lattice oxygen content of the system \cite{SMajumder2019, NBiskup2014}. For SNMO system, O \textit{K} XANES are found to have similar shape and intensity of spectral characters in all the films (data not shown here). This observation confirm no change of lattice oxygen content in different SNMO films. It should be noted that (i) after deposition all SNMO films are cooled under abundant oxygen environment (400 Torr of OPP) to have O stoichiometry. (ii) With increasing oxygen vacancy concentration lattice parameter usually changes \cite{NBiskup2014, SMajumder2019}, which is not observed in the present case. (iii) From chemical valency study, obtained valence state of the constituent elements in SNMO samples are, Sm: 3+, Ni: 55($\pm$1)$ \% $ 2+, 45($\pm$1)$ \% $ 3+, Mn: 54($\pm$2)$ \% $ 4+, 46($\pm$2)$ \% $ 3+. Hence, to maintain overall charge neutrality in Sm$ _{2} $NiMnO$ _{\delta} $, $ \delta $ should be $\sim$5.99. Therefore, we can preclude the possible presence of any major O vacancy defect in the studied SNMO samples. The aforementioned chemical states are distributed throughout the thickness of the films. Observed mixed valency suggests $ Ni^{2+}+Mn^{4+} \longrightarrow Ni^{3+}+Mn^{3+} $ kind of charge disproportionation. Similar charge disproportionation have been observed in many other double perovskite systems \cite{RIDass2003, HZGuok2008, GHJonker1966, JBGoodenough2000}. Note-worthily, with varying disorder concentration we have not observed apparent changes in spectral features of XPS and XANES which, indicates that the core-electronic structure and the charge states of elements present in SNMO films are unbiased to the degree of cation disorder present in the structure. 

\begin{figure*}[t]
\centering
\includegraphics[angle=0,width=0.95\textwidth]{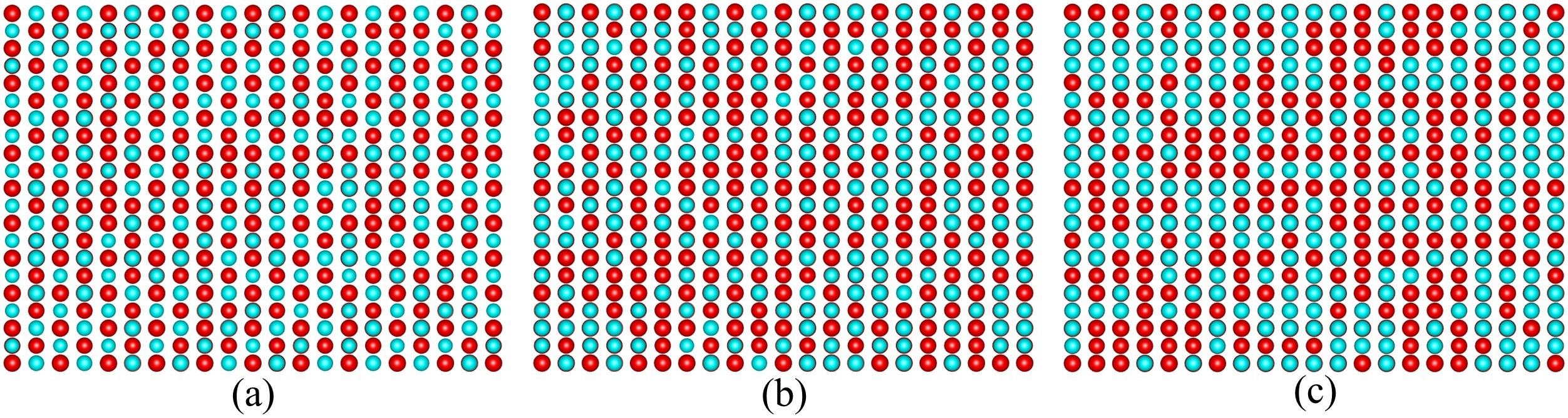}
\caption{Model structures presenting Ni (light blue) and Mn (dark red) cation arrangements at c-b crystallographic plane of SNMO lattice, having different concentration of anti-site occupancy disorders, (a): Q$_{ASD}$ = 5$ \% $, (b): Q$_{ASD}$ = 20$ \% $ and (c): Q$_{ASD}$ = 45$ \% $.}\label{Ni Mn ASD}
\end{figure*}

\subsection{Local structure}\label{Sec Local structure}
EXAFS is a very useful technique to provide valuable information about the local molecular structure, coordination environment of the absorber and short range disorders present in the system. Figures \ref{EXAFS}(a, b) show $ k^{2} $-weighted signal in $k$-space and corresponding modulus of Fourier transforms (FT) of Ni \textit{K}-edge EXAFS spectra respectively, for SNMO thin films. The real and imaginary part of FT are presented in SM \cite{SM}. To investigate the role of B-site cation disorder explicitly, quantitative analysis of $ | \chi (R) |$ has been carried out by model fitting with Sm$ _{2} $NiMnO$ _{6} $ crystal structure (SG: \textit{P2$ _{1} $/n}). Theoretical fitting models are created using crystallographic information obtained from Rietveld analysis of XRD pattern (data not shown here) of polycrystalline SNMO. The unit cell parameters used to generate the theoretical model are tabulated in SM \cite{SM} Table. S1. The fits were confined to R-range of 1 \AA \ $ \leqslant $ R $ \leqslant $ 4 \AA \ and $k$-range of 2.5 \AA$ ^{-1} $ \ $ \leqslant k \leqslant $ 12 \AA$ ^{-1} $. Here within this region, EXAFS spectra originates because of photoelectron scattering from the nearest neighbor octahedral O atoms, the second nearest neighbor Sm atoms and the next neighbor B-site cations (Mn/Ni) connected with Ni core absorber through intermediate O atoms. Depending on the statistical significance, the most relevant single scattering and multiple scattering paths are adopted to model theoretical EXAFS pattern. Along with Ni-O, Ni-Sm and Ni-Mn/Ni single scattering effects, multiple scattering (in forward triangle geometry) effects also have considerable contribution due to the presence of Ni-O-Mn/Ni linkage. 

To define ASD in the fitting model, two types of cells (both having same core as Ni atom) are considered. In first case, all the next B-sites are occupied with Mn atoms (i.e. perfect ordering state) whereas in other case all the next B-sites are filled with Ni atoms (i.e. complete disordered state). The theoretical EXAFS spectra is simulated as a convoluted effect from these two cells \cite{BNRao2016}. The coordination number of Ni-Mn and Ni-Ni bonds are refined to probe the fractional contribution from each cell. To quantify the degree of short range disorder related with the mis-location of B-site cations from their ideal alternating site occupancy, Q$^{XAS} _{ASD} $ parameter is defined as the probability of getting disordered bond configurations. The Q$^{XAS} _{ASD} $ parameter can be expressed as follows \cite{CMeneghini2009},
\begin{eqnarray}\label{EqQASDE}
Q^{XAS}_{ASD} & = & (N_{B-site}-N_{Ni-Mn}) / N_{B-site} = 1 - q  \nonumber\\
q & = & N_{Ni-Mn} / N_{B-site}  \nonumber\\
N_{B-site} & = & N_{Ni-Mn} + N_{Ni-Ni}
\end{eqnarray}
where, N$_{Ni-Mn}$ and N$_{Ni-Ni}$ are the coordination numbers of ordered (Ni-Mn) and disordered (Ni-Ni) bond configurations and N$_{B-site}$ is the total coordination number (N$_{B-site}$ = 6) corresponding to B-site configurations (Ni-Mn/Ni). For all SNMO samples, the total coordination numbers were kept fixed to crystallographic values. Being chemically transferable, amplitude reduction factor (s$_{0}^{2}$) was kept constant at 0.84, a value obtained from the refinement of Ni \textit{K}-edge EXAFS spectra of NiO. Same value of energy shift ($ \Delta $E$ _{0} $) was used for all coordination shells. During refinement cycles, the values corresponding to average coordination distances and mean-square relative displacement (MSRD) factors were refined. Results based on the preliminary trial fittings show that MSRD for Ni-Mn/Ni scattering paths are highly correlated and, therefore same value of $ \sigma^{2} $ can be considered for Ni-Mn/Ni configurations. The goodness of fit was monitored by $ R $ factor which is defined as, 
\begin{equation}\label{EqRfactor}
R = \frac{\sum_{i} [Re(\chi_{d}(R_{i})-\chi_{t}(R_{i}))^{2}+Im(\chi_{d}(R_{i})-\chi_{t}(R_{i}))^{2}]}{\sum_{i} [Re(\chi_{d}(R_{i}))^{2}+Im(\chi_{d}(R_{i}))^{2}]}
\end{equation}
where, $ \chi_{d} $ and $ \chi_{t} $ refer to the experimental and theoretical $ \chi(R) $ values, respectively. Obtained best fit results are superimposed on $ k^{2} . \chi (k)$ and $ | \chi (R) |$ as shown in Figs.\ref{EXAFS}(a, b)). Best fit in Re[$ \chi (R) $] and Im[$ \chi (R) $] are shown in SM \cite{SM}. These model patterns are in a good agreement with the experimentally observed behavior confirmed by the goodness indicator factor (Eq. \ref{EqRfactor}), $ R_{S\_HO} $ $ \sim $ 0.01, $ R_{S\_I} $ $ \sim $ 0.01 and $ R_{S\_HD} $ $ \sim $ 0.009. Refined values of structural parameters are listed in SM \cite{SM} Table S2 and the estimated ASD fractions (Q$_{ASD}^{XAS}$) for the samples are presented in Table \ref{comparison table}. To check the sensitivity of the fitting model over the value of parameter $ q $ (or Q$ _{ASD} $), the best fitted spectra for S$ \_ $HD film with refined value of $ q = 0.53 $ is compared with a model spectra having a fixed (wrong) $ q = 0.90 $ value, which is close to the value of q for S$ \_ $HO case. Clearly, a distinguishable difference is observed in the $ | \chi (R) |$ and Im[$ \chi (R) $] spectra only across the Ni-Mn/Ni coordination shell while comparing the experimentally observed signal and the model generated with $ q $ = 0.90 (fixed at a wrong choice) spectra, as shown in Fig. \ref{EXAFS}(c). This observation confirm that the proposed EXAFS model is appropriate to quantify $ q $ and hence the ASD concentration in SNMO samples. Thus EXAFS analysis unambiguously establishes that local coordination environment around Ni core comprises of both cation ordered and disordered configurations. 

The nature of ASD can be understood on the basis of B-site cation disorder distribution in local atomic structure of SNMO. For a perfectly B-site cation ordered scenario (Q$_{ASD}=0\%$), the alternating octahedral center sites \textit{2c} and \textit{2d} are occupied by Ni and Mn ions respectively with 100$\%$ probability. Whereas, in a completely disordered system, both \textit{2c} and \textit{2d} sites can be occupied by Ni or Mn with 50$\%$ probability for each species (Q$_{ASD}=50\%$), i.e. a random arrangement of B-site cations. A random alloy algorithm on 10$ \times $10$ \times $10 super-cell filled with Ni and Mn atoms on their respective sublattices obeying periodic boundary conditions, is employed to create mis-site defects in SNMO structure. Transition metal ion mislocation at randomly selected sites in the host matrix related model defect structures are simulated with varying fraction of disorder densities as illustrated in Fig.\ref{Ni Mn ASD}(a-c) (for better visualization slice of b-c plane are presented). It is observed that at lower concentration of Q$_{ASD}$, Ni/Mn disordered bonds are distributed homogeneously (Fig.\ref{Ni Mn ASD}(a)), with increasing Q$_{ASD}$ Ni/Mn rich small clusters are formed (Fig.\ref{Ni Mn ASD}(b)) and higher concentration of Q$_{ASD}$ results in partial Ni/Mn segregated patches (Fig.\ref{Ni Mn ASD}(c)) in the lattice. The spatial correlation among disorders depends upon these different nature of ASD distribution patterns.  

\subsection{Magnetic properties}
\begin{figure*}[t]
\centering
\includegraphics[angle=0,width=0.75\textwidth]{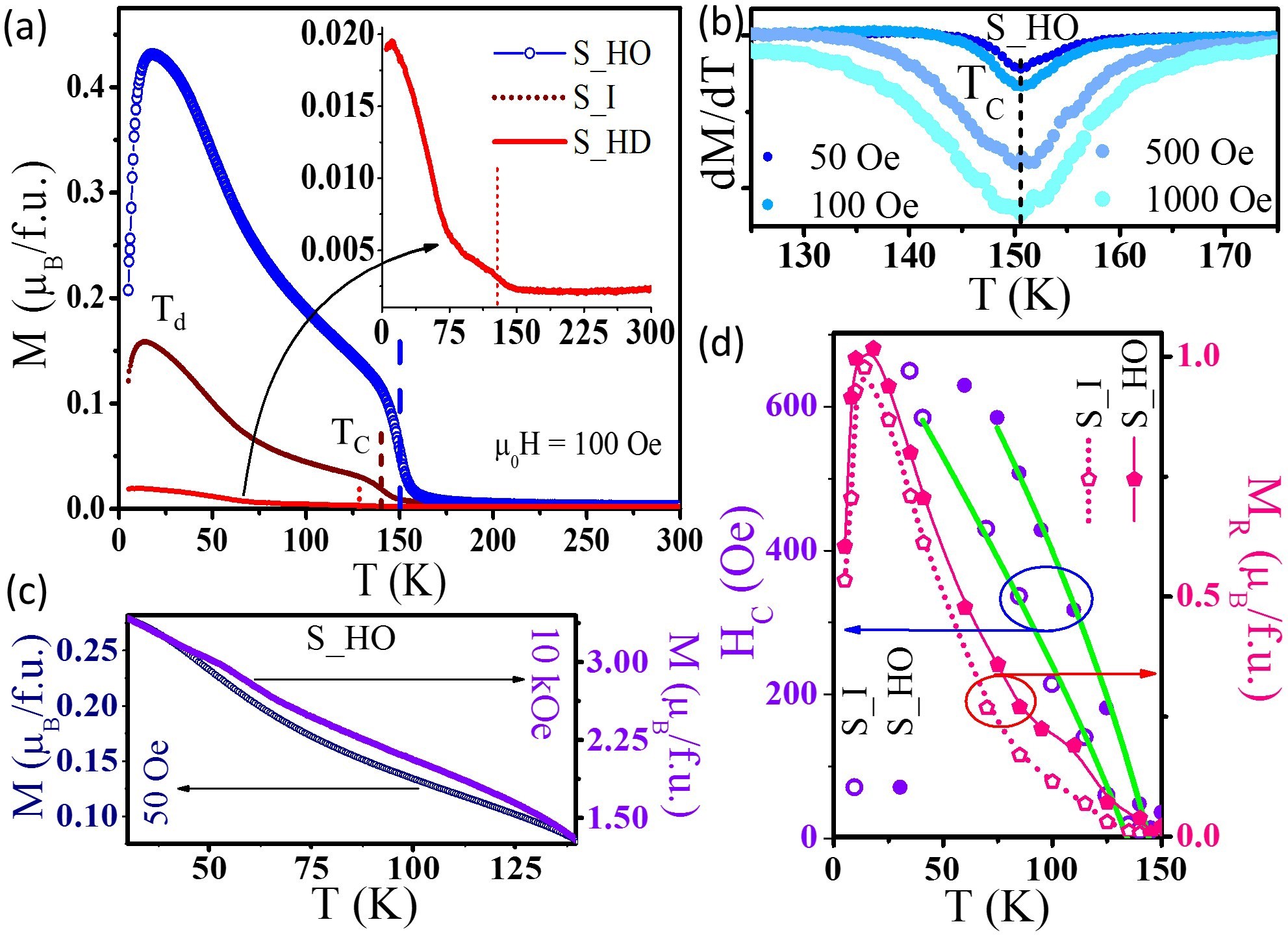}
\caption{(a): Field cooled warming magnetization recorded as a function of temperature M(T) in presence of measuring magnetic field $ \mu_{0} $H=100 Oe. Inset shows enlarged view of M(T) curve for S$ \_ $HD sample. (b): Temperature derivative of magnetization across the paramagnetic to ferromagnetic transition in presence of different applied magnetic fields for S$ \_ $HO sample. (c): M(T) curves for S$ \_ $HO sample, showing effect of different measuring magnetic field values on inverted cusp like trend. (d): Temperature variation of remanence (red pentagons) and coercivity (blue circles are for observed values and solid/green lines are for fits using Eq.(\ref{Eqkron})). The full temperature range of coercivity behavior can be found in SM \cite{SM}. All of these magnetization are measured along in-plane geometry.}\label{MT}
\end{figure*}

We have already observed in Fig.\ref{psd}(b) that the degree of B-site cation ordering have huge impact on saturation magnetization. To further explore the ASD driven modifications in magnetic behavior of SNMO system, we have measured dc magnetization as a function of temperature in presence of different applied magnetic fields and as a function of applied magnetic fields in isothermal conditions. Temperature dependent magnetization M(T) curves of SNMO films in presence of $ \mu_{0} $H=100 Oe of applied magnetic field following typical FCW protocol are presented in Fig.\ref{MT}(a). At first glance, two distinct magnetic transitions are observed, at T=T$ _{C} $ (onset in M(T) highlighted by vertical dashed lines) and at T=T$ _{d} $ (downturn in M(T) at low temperature). 

The transition temperature T$ _{C} $ is estimated from inflection point in first temperature derivative of magnetization dM/dT. T$ _{C} $ is found to be robust against the variation of measuring magnetic field (Fig.\ref{MT}(b)). It is suggested that because of more localized nature of t$ _{2g} $ orbitals only the electrons from half filled e$ _{g} $ orbital of Ni allows to participate in virtual hopping with empty e$ _{g} $ orbitals of Mn, leading to Ni(e$ ^{2}_{g} $)-O-Mn(e$ ^{0}_{g} $) super exchange interaction which is FM via 180$ ^{o} $ linkage \cite{GoodenoughKanamoril195559, PSanyal2017}. Following this prediction, the microscopic magnetic structure of prototype double perovskite La$ _{2} $NiMnO$ _{6} $ (LNMO) is found to be collinear ferromagnetic \cite{NSRogadol2005}. Modified arrott plot analysis of SNMO thin films across T = T$ _{C} $ (not shown here) reveal that the nature of the transition is second order paramagnetic (PM) to FM where mean-field approximation is applicable and the obtained T$ _{C} $ values are within $\pm<$ 2 K with respect to the transition temperature obtained from dM/dT curves. Thus the transition observed at T = T$ _{C} $ in SNMO film is attributed to long range FM ordering of Ni$ ^{2+} $(e$ ^{2}_{g} $)-O-Mn$ ^{4+} $(e$ ^{0}_{g} $) B-site ordered configurations. Furthermore, as the SNMO films show coexisting cation ordered-disordered structures and mixed valence character of both Ni(2+/3+) and Mn(4+/3+) species, a number of additional magnetic interactions are possible in present scenario; (i) B-site cation disorder mediated AFM coupling between Ni$ ^{2+} $(e$ ^{2}_{g} $)-O-Ni$ ^{2+} $(e$ ^{2}_{g} $) and Mn$ ^{4+} $(t$ ^{3}_{2g} $)-O-Mn$ ^{4+} $(t$ ^{3}_{2g} $) pairs due to virtual charge transfer between half filled to half filled orbitals. Such local site disorder related short scale AFM interactions weaken the long range FM ordering. (ii) FM coupling in Ni$ ^{3+} _{LS} $(e$ ^{1}_{g} $)-O-Mn$ ^{3+} _{HS} $(e$ ^{1}_{g} $) (where LS: low spin, HS: high spin) because of vibronic superexchange between singly occupied two fold degenerate e$ _{g} $ orbitals. But, these vibronic superexchange couplings have less stability in comparison with static superexchange (e.g. Ni$ ^{2+} $-O-Mn$ ^{4+} $). (iii) AFM superexchange coupling between Ni$ ^{3+} _{LS} $(e$ ^{1}_{g} $)-O-Ni$ ^{3+} _{LS} $(e$ ^{1}_{g} $) and Mn$ ^{3+} _{HS} $(e$ ^{1}_{g} $)-O-Mn$ ^{3+} _{HS} $(e$ ^{1}_{g} $) disordered pairs. (iv) Similar kind of exchange interaction pathways in between Ni and Mn pairs when Ni$^{3+}$ stabilizes in HS state \cite{FFan2016}. (v) Modified FM exchange interactions due to local distortion at octahedral sites, as both Ni$ ^{3+} $ and Mn$ ^{3+} $ have orbital degeneracy and are John-teller active. Besides the aforementioned exchange interaction pathways, the interfaces of cation ordered and disordered structures can also introduce some additional magnetic interactions (vi): (a) AFM coupling between Ni$ ^{2+} $(e$ ^{2}_{g} $)-O-Ni$ ^{3+} $(e$ ^{1}_{g} $) due to virtual excitation through superexchange in 180$ ^{o} $ geometry, and (b) FM interaction between Mn$ ^{3+} $(e$ ^{1}_{g} $)-O-Mn$ ^{4+} $(e$ ^{0}_{g} $) owing to double exchange interactions. It should be noted that even in the highly ordered sample, there may be AFM interaction at anti-phase boundaries due to presence of Ni$ ^{2+} $-O-Ni$ ^{2+} $, Mn$ ^{4+} $-O-Mn$ ^{4+} $ pairs. These interface effects (vi) are relatively small compare to contribution form cation ordered and disordered regions (i-v). As a result of these coexisting interactions, there is a competition between long range FM interaction and short scale AFM coupling originated from cation ordered and disordered structures respectively, which governs the magnetic behavior in SNMO system. With increasing ASD concentration, contributions from interactions (i) and (iii) increase, causing substantial decrease of overall magnetic moment value and reduction of PM-FM transition temperature T$ _{C} $ (Table \ref{comparison table}). 

\begin{table*}[]
\centering
\begin{tabular}{@{}|c|c|c|c|c|c|c|c|@{}}
\hline 
Sample & T$ _{C} $ (K) & T$ _{d} $ (K) & M$ _{S}^{5 K} $ ($ \mu_{B}/f.u. $) & M$ _{R}^{5 K} $ ($ \mu_{B}/f.u. $) & \multicolumn{2}{c|}{Q$_{ASD}^{M_{S}}$($ \% $) (Eq.(\ref{EqQASDM}))} & Q$_{ASD}^{XAS}$($ \% $) (Eq.(\ref{EqQASDE})) \\
			 &		 &		&	   &	  &	$M_{cal}$ = 5.12 $\mu_{B}$ &	$M_{cal}$ = 5.62 $\mu_{B}$ & \\ \hline
S$ \_ $HO    & 150.1 & 17.7 & 4.96 & 0.61 & 2 & 6 & 5 \\ 
S$ \_ $I     & 140.1 & 14.2 & 3.43 & 0.53 & 17 & 20 & 19 \\ 
S$ \_ $HD    & 128.5 & 10.6 & 0.34 & 0.05 & 47 & 47 & 47 \\ 
\hline
\hline 
\end{tabular}
\caption{Effect of varying anti-site cation disorder in magnetization. Last columns list anti-site disorder fraction estimated from saturation magnetization and extended X-ray absorption fine structure analysis, respectively.}
\label{comparison table}
\end{table*}

Within T$ _{d} < $ T $ <$ T$_{C} $ temperature regime, a broad inverted cusp like trend is prominently observed in highly disordered sample (marked by curved arrow in Fig.\ref{MT}(a)). Similar M(T) behavior is also reported in prototype Pr$ _{2} $NiMnO$ _{6} $ double perovskite system \cite{MPSingh2009}. The inverted cusp attenuates with increasing amount of cation ordered phase fraction while keeping the measuring field fixed (Fig.\ref{MT}(a)) or with increasing measuring field strength for a particular sample having fixed disorder concentration (Fig.\ref{MT}(c)). Here, with increasing the degree of cation ordering, the contribution from short scale AFM coupling decreases, whereas in presence of high measuring magnetic field the long range interaction dominates over diluted short scale couplings. Thus, the inverted cusp like feature can be assigned to the signature of ASDs present in the system. 

\begin{figure*}[t]
\centering
\includegraphics[angle=0,width=0.8\textwidth]{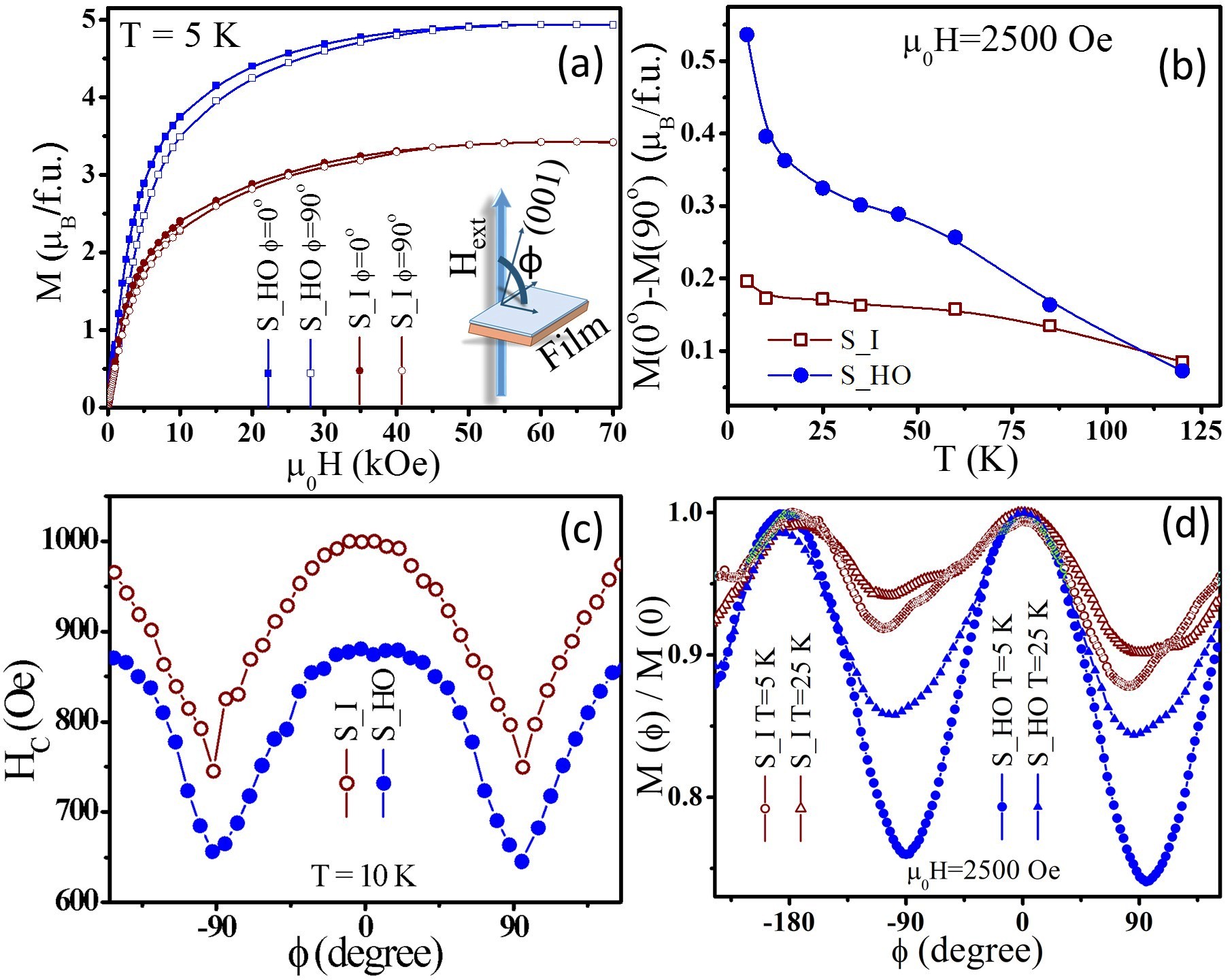}
\caption{(a): Magnetic virgin curves at T=5 K measured along in-plane and out-of-plane geometric configurations. Schematic in inset represents thin film (a-b plane) rotation with respect to external field (H$ _{ext} $) direction for angular dependency of magnetization measurements. (b): Temperature variation of difference in moment obtained from in-plane and out-of-plane geometries. (c): Angular variation of coercivity measured at T=10 K. (d): Magnetic moment as a function of angle measured at T=5 K, 25 K in presence of applied magnetic field of $ \mu_{0} $H=2500 Oe. In all these plots, geometric symbols are for observed data and solid lines are for guide to the eyes.}\label{MA}
\end{figure*}

Magnetic isotherm M(H) recorded at T=5 K (as presented in SM \cite{SM}) shows drastic drop in saturation magnetization with increasing cation disorder density. Considering ideal case, long range ordered FM configurations only, the effective total moment is calculated using, 
\begin{eqnarray}\label{EqMcal}
M_{cal} = [2(\mu_{Sm})^{2} + y(\mu_{Ni^{2+}})^{2} + x(\mu_{Mn^{4+}})^{2} \nonumber\\
+ (1-y)(\mu_{Ni^{3+}})^{2} + (1-x)(\mu_{Mn^{3+}})^{2}]^{1/2} 
\end{eqnarray}
where x and y are fractional concentration for Ni$^{2+}$, Mn$^{4+}$ valence species, respectively obtained from chemical valence state analysis. The estimated $ M_{cal} $ values from Eq.(\ref{EqMcal}) are 5.12 $\mu_{B}$ and 5.62 $\mu_{B}$ corresponding to Ni$^{3+}$ LS and Ni$^{3+}$ HS state, respectively. The order parameter related with ASD configurations Q$ _{ASD}^{M_{S}} $ is defined as \cite{DYang2018, DYang2019}, 
\begin{equation}\label{EqQASDM}
Q_{ASD}^{M_{S}} = 0.5*[1-(M_{S}/M_{cal})]
\end{equation}
which involves the reduction of saturation moment value by disordered AFM pairs originating from Ni species occupying Mn site and vice versa. Estimated values of Q$ _{ASD}^{M_{S}} $ are listed in Table \ref{comparison table}, which are well consistent with EXAFS analysis. Temperature dependency of M(H) behaviors are presented in SM \cite{SM}. The thermal evolution of both remanence (M$ _{R} $) and coercivity (H$ _{C} $) for SNMO samples show non monotonic behavior (Fig.\ref{MT}(d)) indicating evolution of magnetic phases with temperature similar to what we have observed in M(T) measurements. As expected, M$ _{R} $ is observed to decrease with increasing ASD concentration present in the sample (Fig.\ref{MT}(d), Table\ref{comparison table}). The coercive field is the inverse field necessary for reversal of magnetization direction which depends on several mechanisms like nucleation, rotation, propagation motion, pinning and de-pinning of domain walls etc. inherently involved with nature and dimension of the defects present in the system \cite{DGivordl1992}. The temperature dependency of coercivity can be expressed as, \cite{HKronmiilier1988}, 
\begin{equation}\label{Eqkron}
\mu_{0}H_{C} = \lbrace (2 \kappa ) / M_{S} \rbrace \alpha - N_{eff}\mu_{0}M_{S}
\end{equation}
where $ \kappa $ is anisotropy constant, $ \alpha $ is mechanism parameter, M$ _{S} $ is saturation magnetization and N$ _{eff} $ is average dipolar interaction. Mechanism parameter $ \alpha $ depends on microstructure of the magnetic grains which involves anisotropy minimization at the vicinity of the grain boundaries, misalignment of the grains and grain grain exchange interaction. Only in ideal case for magnetically decoupled, perfectly aligned grains, the $ \alpha $ = 1 condition holds. In practical situations, all such contributions involves in $ \alpha $ are less than unity. The effective anisotropy energy of SNMO films are estimated using the following equation,
\begin{equation}\label{Eqkani}
\kappa_{eff} = \int_{0}^{\mu_{0}H_{S}}M(H)_{in} dH - \int_{0}^{\mu_{0}H_{S}}M(H)_{out} dH
\end{equation}
where H$ _{S} $ is saturation field, in and out correspond to in-plane and out-of-plane geometric configurations of magnetization with respect to applied magnetic field direction, respectively, as presented in Fig. \ref{MA}(a). The obtained values of $\kappa_{eff}$ are as follows: $\kappa_{eff}$(S$\_$HO) = 6.472 $ \times $ 10$ ^{5} $ erg/cm$ ^{3} $ and $\kappa_{eff}$(S$\_$I) = 2.696 $ \times $ 10$ ^{5} $ erg/cm$ ^{3} $. The calculated variation of coercivity in T$ _{d} < $ T $ <$ T$_{C} $ regime is presented in Fig.\ref{MT}(d), while the full temperature range behavior can be found in SM \cite{SM}. Fitting of experimental data by calculated pattern yields the following values of the parameters: $\alpha$(S$\_$HO) = 0.044(1), N$_{eff}$(S$\_$HO) = 0.215(3) and $\alpha$(S$\_$I) = 0.067(4), N$_{eff}$(S$\_$HO) = 0.194(5). Such small values of $\alpha$ indicate dominating pinning mechanism over nucleation in magnetization reversal process with narrow heterogeneity of the magnetic phase present in the samples \cite{DGivordl1992, HKronmiilier1988}. For narrow heterogeneity the pinning efficiency varies as $ \sim $ d/$ \delta $ where d and $ \delta $ represent widths of the defect and domain wall, respectively \cite{DGivordl1992, HKronmiilier1988}. Larger value of $ \alpha $ for S$ \_ $I in comparison with S$ \_ $HO, implies larger dimension of defects in S$ \_ $I than that of S$ \_ $HO, which is due to convoluted effect of more mislocated site disorder present in S$ \_ $I. 

The value of in-plane M$ _{R} $ is higher than the out-of-plane M$ _{R} $, indicating that it is easier to magnetize the films along in-plane direction and therefore, magnetic easy axis lies in the a-b plane. The virgin magnetization curves recorded at T=5 K along in-plane and out-of-plane directions for SNMO films are displayed in Fig. \ref{MA}(a). In order to further elucidate the symmetry distribution of anisotropy axes, we have measured magnetization as a function of angle $ \phi $ between a-b plane (easy axis) of the films and applied field (H$ _{ext} $) direction (as illustrated by schematic shown in Fig. \ref{MA}(a)). The difference  in magnetic moment values $ \delta $M along $ \phi $ = 0$ ^{o} $ and $ \phi $ = 90$ ^{o} $, measured as a function of temperature below T$ _{C} $ region are presented in Fig.\ref{MA}(b). The angular variation of H$ _{C} $ curves (Fig.\ref{MA}(c)) reveal two fold symmetry as the two minima are occuring within -155$ ^{0} \leqslant \phi\leqslant $ +155$ ^{o} $ range. Similarly, in M($ \phi $) measurements (Fig.\ref{MA}(d)), the two fold anisotropy distribution is observed in 360$ ^{o} $ angular scans. Thus the observed angular dependency of magnetization suggests predominant uniaxial nature of anisotropy in the films. One should notice that the angular difference between the easy direction and the immediate hard direction or the $\delta \phi$ width representing half periodicity, is not exactly 90$ ^{o} $ (95.1$ ^{o} $ for S$ \_ $HO and 79.9$ ^{o} $ for S$ \_ $I) in these samples. This observation suggests the presence of additional weak anisotropy contribution such as biaxial term along with dominating uniaxial contribution. The thermal evolution of angle dependent magnetization behaviors are displayed in SM \cite{SM}. Comparison between SNMO samples having different extent of disorder densities reveals that with increasing cation disorder in the system, deviation from uniaxial behavior increases and anisotropy reduces considerably. Observed behavior is due to increase of disorder related AFM phase which weakens the long range FM interaction in SNMO. 

\begin{figure*}[t]
\centering
\includegraphics[angle=0,width=0.8\textwidth]{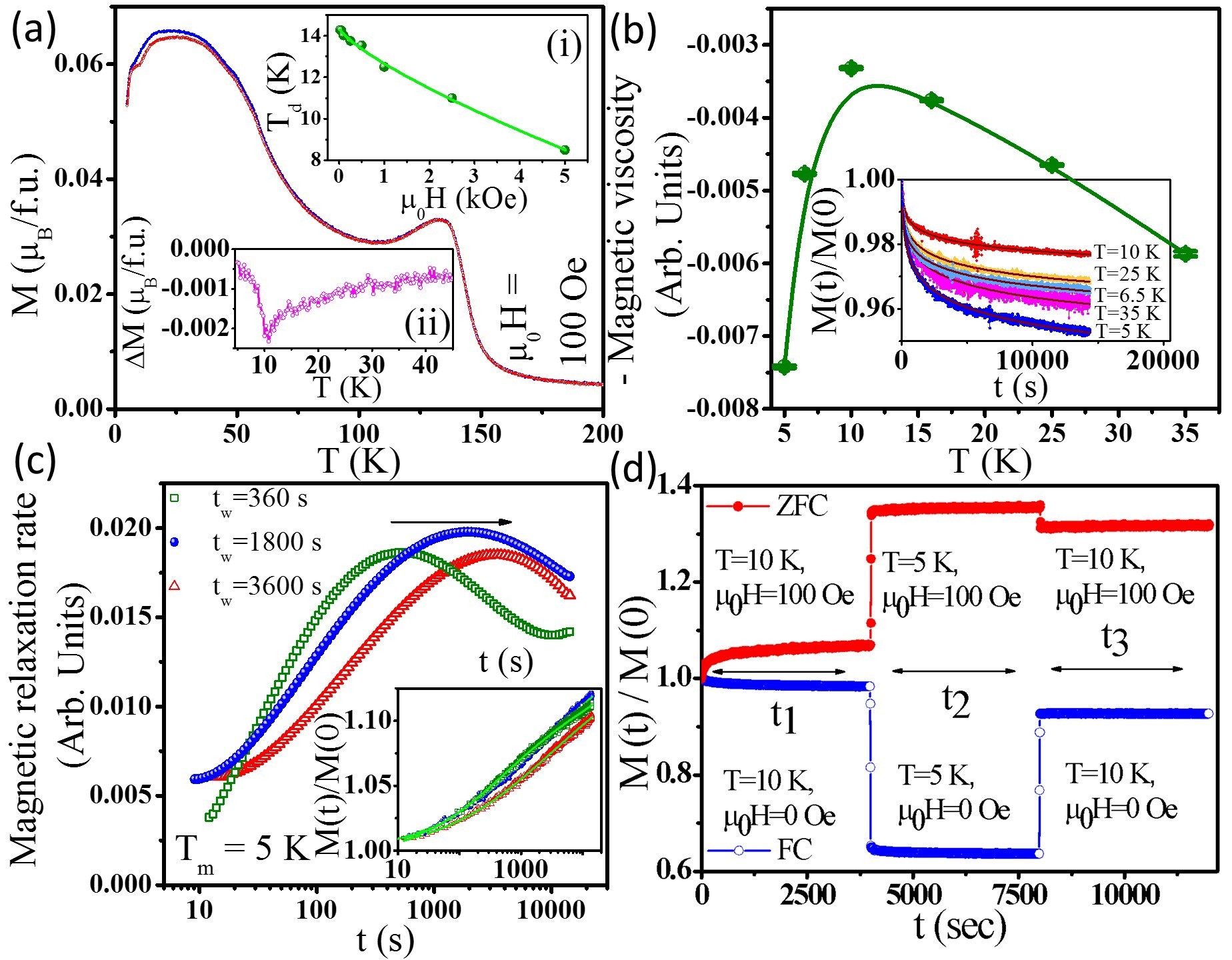}
\caption{(a): Comparison between zero field cooled reference (M$ _{ZFC}^{REF} $) curve (blue) and single stop wait (M$ _{ZFC}^{SSW} $) curve (red) measured in presence of $ \mu_{0} $H=100 Oe applied magnetic field. Inset (ii) shows temperature variation of $ \Delta $M = M$ _{ZFC}^{SSW} $ - M$ _{ZFC}^{REF} $. Inset (i) presents transition temperature T$ _{d} $ as a function of measuring magnetic field $\mu_{0}$H, dots (dark green): observed behavior, solid line (light green): fitted curve with Eq.(\ref{EQHT}). (b): Magnetic viscosity as a function temperature, obtained by fitting the fieled cooled magnetization relaxation (shown in the Inset) behavior with Eq.(\ref{EQMt}). (c) Relaxation rates calculated from relaxation of zero field cooled magnetization (shown in the Inset) with different wait times t$ _{w} $, using Eq.(\ref{EQSt}). (d): Zero field cooled (red) and field cooled (blue) relaxation with intermediate negative thermal cycling. All of these magnetization are measured along in-plane geometry.}\label{Mtime}
\end{figure*}

Some prototypical members of RNMO family show downturn in magnetization below T$_{d}$ \cite{DCKakarla2014, RYadava2015}, although the origin of the low temperature magnetic transition at T$_{d}$ is still elusive. In case of LNMO system, the magnetic phase transition occurring at T = T$_{d}$ is attributed to the possible formation of cation disorder mediated reentrant spin glass (SG) or reentrant ferromagnet like phase \cite{PSanyal2017, DChoudhury2012}. To investigate the magnetization behavior below T = T$ _{d} $ in SNMO system, firstly we have selected S$ \_ $I film for detailed study, which have approximately moderate phase fraction of both ordered and disordered structures. Afterwards, we will analyze the effect of different ASD concentration by comparing the results corresponding to all the three samples. On reducing the temperature below T$ _{d} $, we have observed that both ZFC and FCW magnetization curves decrease sharply under low ($ \mu_{0} $H$ \ll $1.5 T) applied field. The detailed results are presented in SM \cite{SM}. Usually the M$ _{FC} $ becomes a constant or decreases very slowly with respect to the temperature variation in case of a SG phase, while it increases with decreasing temperature for superparamagnets \cite{MSuzuki2009}. We have observed that the transition temperature T$ _{d} $ depends on the magnitude of measuring dc magnetic field $ \mu_{0} $H, as shown in Fig.\ref{Mtime}(a):Inset(i) and the variation in H-T space can be described  by \cite{MSuzuki2009}, 
\begin{equation}\label{EQHT}
T_{d}(H) = T_{d}(0) \left\lbrace 1 - \left( H/H_{a} \right)  ^{p} \right\rbrace 
\end{equation}
where, T$ _{d} $(0) estimates the transition temperature at zero field, H$_{a}$ is the field amplitude and $ p $ is the exponent. Similar evolution of transition temperature with external filed strength forming a critical line in H-T space is reported in SG \cite{MSuzuki2009} as well as in uniform ferromagnet system with cubic anisotropy \cite{KHFischer1987}. H-T space critical line in SG system defines border line for SG phase that is upon crossing this line relaxation time and correlation length diverge. In ferromagnetic system with cubic anisotropy H-T space critical line defines border line for metastable and irreversible phase. In case of SG phase, exponent $ p $ characterizes the type of phase transition. For instance $ p $ = 2/3, namely de Almeida-Thouless (AT) line \cite{deAlmeida1978} is for Ising type SG and $ p $ = 2, namely Gabay-Thouless (GT) line \cite{MGabay1981} is for freezing of transverse spin components.  Least-squares fitting of the observed data in H-T space yields T$ _{d} $ = 14.3$ \pm $0.1 K, H$ _{a} $ = 16.22$ \pm $1.05 kOe, $ p $ = 0.76$ \pm $0.07, which indicates close to AT-line behavior in H-T phase diagram. The field dependent downturn transition at T$ _{d} $ is further investigated by exploring the dynamical response of dc-magnetization. Characteristic ZFC memory effect is measured following the typical single stop and wait procedure (SSW) in ZFC protocol (Fig.\ref{Mtime}(a)). For this measurement, at first a reference ZFC curve M$ _{ZFC}^{REF} $ is recorded in standard ZFC mode in presence of magnetic field $ \mu_{0} $H = 100 Oe. In next turn during zero magnetic field cooling, the sample was cooled from T$>$T$ _{C} $ down to a stop temperature T$ _{s} $ = 10 K, aged at T = T$ _{s} $ for a wait time t$ _{w} $ = 10$ ^{4} $s, then cooled down to 5 K and, the magnetization M$ _{ZFC}^{SSW} $ is again recorded under applied magnetic field $ \mu_{0} $H = 100 Oe. Temperature dependency of difference curve defined as $ \Delta $M = M$ _{ZFC}^{SSW} $ - M$ _{ZFC}^{REF} $ (presented in Fig.\ref{Mtime}(a:Inset(ii)) shows distinct dip just near T$ _{s}$ at T$\simeq$10.7 K. This observation reveals that system memories its cooling  history. During isothermal aging at T = T$ _{s} $ for t$ _{w} $ duration, the spins spontaneously arrange towards equilibrium state through the growth of equilibrium domains and on further cooling below T$ _{s} $ they are frozen. The spins retrieve the memory on reheating in this process \cite{MSuzuki2009, SSahoo2003}. 

To further explore the non-equilibrium dynamics of system, relaxation measurements under small field perturbation are carried out. To record the relaxation of FC (ZFC) magnetization, system is cooled in presence of magnetic field $ \mu_{0} $H = 100 Oe (zero Oe) from T = 300 K down to measuring temperature T$ _{m} $ and then after an isothermal wait time of t$ _{w} $ applied field is turned off (field $ \mu_{0} $H = 100 Oe is applied) and isothermal time evolution of magnetic moment is measured. The time decay of FC remanent magnetization with t$ _{w} $ = 0 s across T$ _{d} $ (presented in Fig.\ref{Mtime}(b:Inset)) shows non-saturating  behavior within the time scale involved in the measurements. The observed time evolution of magnetization M(t) can be described by logarithmic decay function \cite{LSLakshmi2011}, 
\begin{equation}\label{EQMt}
M(t) = M_{0} [1 - S \ ln(1+t/t_{0})] 
\end{equation}
where M$ _{0} $ is initial remanent magnetization, S $ \equiv $ S(T, H) is magnetic viscosity which characterizes the relaxation mechanism and depends on the material, t$ _{0} $ is reference time involved with measurement process. The random or mixed spin interactions or time evolved activation mediated energy barrier distribution results in this kind of logarithmic relaxation behavior \cite{LSLakshmi2011}. In general, the FC relaxation of magnetic moment decreases monotonically with increasing measuring temperature due to enhancement in thermal fluctuations \cite{YSun2003}. However, it should be noted that the observed temperature variation of FC relaxation process in our measurements shows non-monotonic behavior. The estimated values of magnetic viscosity attain an inflection point at T$ \simeq $10 K (Fig.\ref{Mtime}(b)), which is a signature of spin glass or frozen ferro-fluid phases \cite{LSLakshmi2011}. Such viscosity behavior indicates a trade-off between freezing of spins due to competing exchange interactions and thermal activation of frozen spins with increasing temperature \cite{LSLakshmi2011}. Wait time t$ _{w} $ dependency (aging effect) is observed in ZFC relaxation measurements recorded at T = 5 K ($ < $T$ _{d} $)(presented in Fig.\ref{Mtime}(c:Inset)). The relaxation rate defined as,
\begin{equation}\label{EQSt}
S(t) = (1/H)(\partial M(t)/\partial lnt)
\end{equation}
shows characteristic maxima around t = t$ _{w} $ (Fig.\ref{Mtime}(c)) revealing the presence of aging effect. Similar aging behavior is observed in magnetically disordered and frustrated systems including SG and chaotic reentrant ferromagnets \cite{LSLakshmi2011, MSasaki2005}. After quenching the system below T$ _{d} $ (T$ _{m} <$T$ _{d} $) during the aged time t$ _{w} $, as the system is left unperturbed (no change in T or H), growth of equilibrium domains start and the observed aging signature at around t $ \approx $ t$ _{w} $ is related to the dynamic response of system originating because of crossover from quasi-equilibrium regime (t$ \leq $t$ _{w} $) to non-equilibrium regime (t$ \geq $t$ _{w} $) \cite{LSLakshmi2011, SSahoo2003}. ZFC and FC relaxation experiments with intermediate negative thermal cycling ($ \Delta $T = 5 K), below T = T$ _{d} $, are also examined (Fig.\ref{Mtime}(d)). Generally, relaxation before and after negative thermal cyclings are expected to follow continuous trend without having shift in magnetization value for SG \cite{YSun2003}. Observed discontinuity in relaxation behavior at t$ _{1} $ and t$ _{3} $ time scales for $ \Delta $T = 5 K, indicates the presence of more than one coexisting relaxation process. 

Observed sharp decrease of moment in temperature dependent magnetization M(T) measurements below T=T$ _{d} $ as presented in SM \cite{SM} Fig. S10(a), is possibly due to opposite alignment of Sm PM moments with respect to Ni-Mn sublattice moments polarized in presence of internal exchange field. Similar internal field polarization of rear earth moments antiparallel to transition metal moments has been reported in some perovskite as well double perovskite structures \cite{MTripathi2019, JSBenitez2011, SPal2019}. As shown in SM \cite{SM} Fig. S10(d), with increasing the magnitude of applied magnetic field ($ \mu_{0} $H $ < $ 1.5 T), the downturn behavior vanishes (at $ \mu_{0} $H $ \sim $ 1.5 T). This indicates a balance of internal field at Sm sites by external applied field (hence H$_{int} \simeq $ 1.5 T). A further increase in magnetic field ($ \mu_{0} $H $ > $ 1.5 T) transforms the downturn into an upturn behavior. In disordered magnetic systems, presence of AFM interactions in FM host matrix causes spin frustration \cite{KJonason1996}. Here, below T$ _{d} $ the inter-competing interactions from FM and AFM spin arrangements between Ni-Mn sublattice and Sm-(Ni/Mn) sublattice respectively, may cause similar frustrated magnetic states in SNMO. Consequently, at low temperature values, instead of single global minima (which is generally observed for equilibrium states), multiple local minima pockets with finite barrier heights are formed in free energy landscapes and metastable nature in dynamical response of magnetization is observed in SNMO thin film system. 

Similar signatures of non-equilibrium magnetic state are observed in the measurements performed for all the SNMO samples. With increasing B-site cation disorder in the system, the transition T$ _{d} $ shifts towards lower temperature value (Table \ref{comparison table}) and it's sharpness reduces. Here, we argue that the ASD related short scale AFM interactions are not the dominating cause of such magnetic frustration in SNMO at low temperature (T $ \leqslant $ T$ _{d} $) otherwise non-equilibrium dynamical behavior should be observed above T$ _{d} $ as well, where both order disorder structures related FM-AFM interactions coexist, and also, with increasing disorder or in other words, with increasing ASD AFM pairs, T$ _{d} $ should show a rise in temperature, which is contradictory to our observations. The polarization of Sm PM moments depends on long range ordering of Ni-Mn sublattices. Introduction of B-site cation disorder weakens the Ni-Mn long range interaction and hence the internal field strength reduces, which affect the downturn transition by shifting it towards lower temperature values.

From the aforementioned discussions, it is transpired that the varying level of cation occupancy defects in SNMO system has immense impact on magnetization behavior. Apart from this, in all the SNMO samples studied here, the presence of surface island and edge dislocation defects are found, which may be the possible way to relax the misfit strain from the substrate. Observed similar RHEED patterns indicate similar island-layer surface creation in all SNMO thin films. The similar growth modes as well as similar strain relaxation process in the studied films point out that cation distribution is not altered by the strain relaxation process. In rocking curve, extremely narrow and similar FWHM values for all the films, which is found comparable to single crystal STO substrate, suggest very low concentration of dislocation defects in SNMO samples and their similar distribution. These misfit defects may introduce local tilt in transition metal octahedra leading to possible modification in magnetic exchange by changing bond angle around defect site. It should be noted here that such defect mediated modification of magnetic exchange interaction must have low impact (due to low density) and similar extent (due to similar distribution) in all SNMO samples. Grain boundary crack is another possible defect which may influence the measured magnetic properties of the system by introducing moment pinning centers. However, since all the films are epitaxial, have similar smaller FWHM values in rocking curves and narrow heterogeneity in the magnetic phase as observed from magnetic coercivity analysis, the contribution of the grain boundaries is minimal and comparable in all the films. Therefore, in the present scenario, among all possible defects, ASD plays the most dominating character in governing the magnetic state of the SNMO double perovskite system and the magnetic observables can be tuned by engineering ASD in the host matrix.

\section{CONCLUSION}
In summary, we have mapped the consequences of anti-site cation disorders over the microscopic and macroscopic observables determining the electronic and magnetic properties of epitaxial Sm$ _{2} $NiMnO$ _{6} $ thin film system. The level of cation ordering is found to be sensitive against fine tuning of growth parameters used for the fabrication process and we have presented corresponding phase stability diagram which provides the recipe to engineer anti-site disorder concentration. Local probe extended X-ray absorption fine structure analysis implementing a proper model have been employed to quantify the anti-site disorder fractions in the samples grown under distinct deposition conditions. Density of cation disorders are again evaluated on a global scale using bulk magnetometric measurements and the results are well consistent with estimations of local structure studies. Moreover, from structural simulations we have observed that with increasing concentration, disorder distribution transforms from homogeneous to partial segregation type. The electronic structure of these samples exhibits mixed valence nature of both Ni and Mn species, confirming the $ Ni^{2+}+Mn^{4+} \longrightarrow Ni^{3+}+Mn^{3+} $ kind of charge disproportionation occurring in the system, which is insensitive to the degree of cation ordering. The introduction of disordered Ni-O-Ni and Mn-O-Mn bonds in the structure causes local AFM correlations in the background of ordered Ni-O-Mn FM interactions, and as a consequence, the magnetic behavior of SNMO comprise of coexisting long range FM ordering and short scale AFM interactions. The mutual competing nature of these two phases depends on the anti-site disorder fraction, and leads to the observed decrease in FM transition temperature T$ _{C} $, drastic drop in saturation moment value M$ _{S} $, reduction of remanence magnetization M$ _{R} $ and anisotropy energy $ \kappa $ with increasing anti-site disorder density. We suggest that the field dependent inverted cusp like trends in M(T) are the direct signatures of the anti-site disorders, present in the system. The non-monotonic nature of coercive field H$ _{C} $(T) and remanence magnetization M$ _{R} $(T) originate because of distinct temperature dependent magnetic contributions from different interaction paths. The magnetization reversal process is more likely governed by the pinning mechanism in comparison to the nucleation process having narrow heterogeneity of magnetic phase and dimension of defects involving the pinning process increases with an increasing fraction of anti-site disorder. Angular dependency of magnetization reveals that the presence of anti-site disorders introduce additional bi-axial anisotropy contribution in the predominant uniaxial anisotropy character of the system. Below T$ _{d} $ transition, observed magnetization behavior is attributed to the frustration due to inter competing magnetic interactions from polarized Sm PM moments and Ni-Mn sublattice FM moments, which generates the landscape of multiple non-uniform free energy barriers and drives the system into metastable state. Interplay between internal field and magnitude of applied magnetic field governs the low temperature (across T$ _{d} $) M(T) behavior, which can be transformed from downturn to upturn by varying measuring field strength. The presence of anti-site disorder breaks the long range FM chains and as a consequence significantly decreases the internal magnetic field acting on Sm PM moments and hence, with increasing anti-site disorder the T$ _{d} $ transition moves to lower temperature. Thus, the presence of Ni/Mn mis-site defects in SNMO thin films have huge bearing on magnetic properties of the system, which can be tailored by proper control of synthesis conditions. The present study will in general, help to correlate the modification in the functional properties of a double perovskite system mediated by cation disorders in the host matrix.  
\\
\section*{ACKNOWLEDGMENTS}
Thanks to Elettra-Sincrotrone, Italy and Indus Synchrotron RRCAT, India for giving access to experimental facilities. Authors gratefully acknowledge the Department of Science and Technology, Government of India, Indian Institute of Science, Italian Government and Elettra for providing financial support through Indo-Italian Program of Cooperation (No. INT/ITALY/P-22/2016 (SP)) to perform experiments at Elettra-Sincrotrone. Thanks to Dr. V. R. Reddy (UGC DAE CSR, India) for RSM, Mr. A. Gome for X-ray rocking and Mr. A. Wadikar, Mr. S. Karwal, Mr. M. Kumar for their technical help in measurements performed at RRCAT and CSR. S.M. thanks Mr. Akash Surampalli (UGC DAE CSR) for helpful discussions.

\bibliography{}

\begin{thebibliography}{999}

\bibitem{RNoriega2013}
R. Noriega, J. Rivnay, K. Vandewal,  F. P. V. Koch, N. Stingelin, P. Smith, M. F. Toney and A. Salleo, 
Nat. Mater \textbf{12}, 1038-1044 (2013).

\bibitem{AHusmann1996}
A. Husmann, D. S. Jin, Y. V. Zastavker, T. F. Rosenbaum, X. Yao, J. M. Honig, 
Science \textbf{274}, 5294, 1874-1876 (1996).

\bibitem{ABalandin2011}
A. Balandin, 
Nat. Mater \textbf{10}, 569-581 (2011).

\bibitem{JBowles2013}
J. A. Bowles, M. J. Jackson, T. S. Berqu\'{o}, P. A. S{\o}lheid and J. S. Gee, 
Nat. Commun. \textbf{4}, 1916 (2013).

\bibitem{MTAnderson1993}
M. T. Anderson, K. B. Greenwood, G. A. Taylor, K. R. Poeppelmeier, 
Prog. in Solid St. Chem, \textbf{22}, 3, 197-233 (1993).

\bibitem{MGHernandez2001}
M. Garc\'{i}a-Hern\'{a}ndez, J. L. Mart\'{i}nez, M. J. Mart\'{i}nez-Lope, M. T. Casais, and J. A. Alonso, 
Phys. Rev. Lett. \textbf{86}, 2443 (2001).

\bibitem{DDSarma2001}
D. D. Sarma, S. Ray, K. Tanaka, M. Kobayashi, A. Fujimori, P. Sanyal, H. R. Krishnamurthy, and C. Dasgupta, 
Phys. Rev. Lett. \textbf{98}, 157205 (2007).

\bibitem{DChoudhury2012}
D. Choudhury, P. Mandal, R. Mathieu, A. Hazarika, S. Rajan, A. Sundaresan, U. V. Waghmare, R. Knut, O. Karis, P. Nordblad and D. D. Sarma, 
Phys. Rev. Lett \textbf{108}, 127201 (2012).

\bibitem{NSRogadol2005}
N. S. Rogado, J. Li  A. W. Sleight, M. A. Subramanian, 
Adv. Mater., \textbf{17}: 2225-2227 (2005).

\bibitem{HJZhaonat2014}
H. J. Zhao, W. Ren, Y. Yang, J. \'{I}\~{n}iguez, X. M. Chen and L. Bellaiche, 
Nat. Commun. \textbf{5}, 4021 (2014). 

\bibitem{HJZhaoprb2014}
H. J. Zhao, X. Q. Liu, X. M. Chen, and L. Bellaiche, 
Phys. Rev. B \textbf{90}, 195147 (2014).

\bibitem{GoodenoughKanamoril195559}
(i) J. B. Goodenough, 
Phys. Rev. \textbf{100}, 564 (1955); 
(ii) J. Kanamori, 
J. Phys. Chem. Solids \textbf{10}, 87 (1959).

\bibitem{CSohn2019}
C. Sohn, E. Skoropata, Y. Choi, X. Gao, A. Rastogi, A. Huon, M. A. McGuire, L. Nuckols, Y. Zhang, J. W. Freeland, D. Haskel and H. N. Lee, 
Adv. Mater. \textbf{31}, 1805389 (2019).

\bibitem{JFrantt2019}
J. Frantti, Y. Fujioka, C. Rouleau, A. Steffen, A. Puretzky, N. Lavrik, I. N. Ivanov and H. M. Meyer, 
J. Phys. Chem. C, \textbf{123}, 32, 19970-19978 (2019).

\bibitem{YFujioka2019} 
Y. Fujioka, J. Frantti, C. Rouleau, A. Puretzky, Z. Gai, N. Lavrik, A. Herklotz, I. Ivanov and H. Meyer, 
Annalen Der Physik \textbf{531}, 1900299 (2019).

\bibitem{YShiomi2014}
Y. Shiomi and E. Saitoh, 
Phys. Rev. Lett \textbf{113}, 266602 (2014).

\bibitem{MAzuma2005}Masaki Azuma, 
M. Azuma, K. Takata, T. Saito, S. Ishiwata, Y. Shimakawa and M. Takano, 
J. Am. Chem. Soc., \textbf{127}, 24, 8889-8892 (2005).

\bibitem{JSu2015} 
J. Su, Z. Z. Yang, X. M. Lu, J. T. Zhang, L. Gu, C. J. Lu, Q. C. Li, J.-M. Liu and J. S. Zhu, 
ACS Appl. Mater. Interfaces \textbf{7}, 24, 13260-13265 (2015).

\bibitem{KHJBuschow1974}
K.H.J. Buschow, A.M. van Diepen, H.W. de Wijn, 
Solid St. Commun., \textbf{15}, 5 903-906 (1974) and references therein.

\bibitem{BRavel2005}
B. Ravel and M. Newville, 
J. Synchrotron Radiat. \textbf{12}, 537 (2005).

\bibitem{MNewville1993}
M. Newville, P. L\ifmmode \bar{\imath}\else \={\i}\fi{}vi\ifmmode \mbox{\c{n}}\else \c{n}\fi{}\ifmmode \check{s}\else \v{s}\fi{}, Y. Yacoby, J. J. Rehr and E. A. Stern, 
Phys. Rev. B \textbf{47}, 14126 (1993).

\bibitem{BRavel2001}
B. Ravel, 
J. Synchrotron Radiat. \textbf{8}, 314 (2001).

\bibitem{JJRehr200009}
(i) J. J. Rehr, R. C. Albers, 
Rev. Mod. Phys. \textbf{72}, 621 (2000); 
(ii) J. J. Rehr, J. J. Kas, M. P. Prange, A. P.Sorini, Y. Takimoto, F. Vila, 
C. R. Phys. \textbf{10}, 548 (2009).

\bibitem{SM}
See Supplemental Material at `' for a description of crystallographic information used in EXAFS modeling, structural parameters obtained form EXAFS analysis, supporting results for structural characterization and various magnetometric measurements.

\bibitem{RIDass2003}
R. I. Dass, J.-Q. Yan, and J. B. Goodenough, 
Phys. Rev. B \textbf{68}, 064415 (2003).

\bibitem{DYang2018}
D. Yang, W. Wang, T. Yang, G. I. Lampronti, H. Ye, L. Wu1, Q. Yu and S. Lu, 
APL Mater. \textbf{6}, 066102 (2018). 

\bibitem{DYang2019}
D. Yang, G. I. Lampronti, C. R. S. Haines and M. A. Carpenter, 
Phys Rev. B \textbf{100}, 014304 (2019).

\bibitem{XYang2017}
X. Yang, H. Wuc, X. Wang, Y. Luod, L. Li, 
J. Alloys Compd. \textbf{723}, 930-935 (2017).

\bibitem{SEKaczmarek2017}
S. E. Kaczmarek, and B. P. Thornton, 
Chem. Geol. \textbf{468}, 32-41 (2017).

\bibitem{HZGuok2008}
H. Z. Guo, J. Burgess, E. Ada, S. Street, A. Gupta, M. N. Iliev, A. J. Kellock, C. Magen, M. Varela and S. J. Pennycook, 
Phys. Rev. B \textbf{77}, 174423 (2008).

\bibitem{JPMaria1998}
J-P. Maria, S. Trolier-McKinstry and D. G. Schlom, 
J. Appl. Phys. \textbf{83}, 4373 (1998).

\bibitem{AFluri2018}
A. Fluri, C. W. Schneider, D. Pergolesi, 
In Metal Oxides, Metal Oxide-Based Thin Film Structures, Elsevier, Pages 109-132, ISBN 9780128111666, (2018).

\bibitem{PGay1953}
P Gay, P.B Hirsch, A Kelly, 
Acta Metall. \textbf{1}, 3 315-319 (1953).

\bibitem{NMannella2008}
N. Mannella, C. H. Booth, A. Rosenhahn, B. C. Sell, A. Nambu, S. Marchesini, B. S. Mun, S.-H. Yang, M. Watanabe, K. Ibrahim, E. Arenholz, A. Young, J. Guo, Y. Tomioka and C. S. Fadley, 
Phys. Rev. B \textbf{77}, 125134 (2008).

\bibitem{LSangaletti1995}
L. Sangaletti, L. E. Depero, P. S. Bagus, and F. Parmigiani, 
Chem. Phys. Lett. \textbf{245}, 463 (1995).

\bibitem{VRGalakhov2002}
V. R. Galakhov, M. Demeter, S. Bartkowski, M. Neumann, N. A. Ovechkina, E. Z. Kurmaev, N. I. Lobachevskaya, Y. M. Mukowskii, J. Mitchell, and D. L. Ederer, 
Phys. Rev. B \textbf{65}, 113102 (2002).

\bibitem{APGrosvenor2006}
A. P. Grosvenor, M. C. Biesinger, R. St.C. Smart, N. S. McIntyre, 
Surface Science \textbf{600} 1771-1779 (2006) and references therein.

\bibitem{AEBocquet199295}
A. E. Bocquet, T. Mizokawa, T. Saitoh, H. Namatame and A. Fujimori, 
Phys. Rev. B \textbf{46} 3771-3784 (1992-I).

\bibitem{FBridges2001}
F. Bridges, C. H. Booth, M. Anderson, G. H. Kwei, J. J. Neumeier, J. Snyder, J. Mitchell, J. S. Gardner, and E. Brosha, 
Phys. Rev. B \textbf{63} 214405 (2001).

\bibitem{AHdeVries2003}
A. H. de Vries, L. Hozoi, R. Broer, 
International Journal of Quantum Chemistry \textbf{91} 57-61 (2003).

\bibitem{deGroot1989}
F. M. F. de Groot, M. Grioni, J. C. Fuggle, J. Ghijsen, G. A. Sawatzky and H. Petersen, 
Phys. Rev. B \textbf{40}, 8 5715-5723 (1989-I).

\bibitem{SMajumder2019}
S. Majumder, P. Basera, M. Tripathi, R. J. Choudhary, S. Bhattacharya, K. Bapna and D. M. Phase, 
J. Phys.: Condens. Matter \textbf{31}, 205001 (2019).

\bibitem{NBiskup2014}
N. Bi\v{s}kup, J. Salafranca, V. Mehta, M. P. Oxley, Y. Suzuki,
S. J. Pennycook, S. T. Pantelides and M. Varela, 
Phys. Rev. Lett. \textbf{112}, 087202 (2014).

\bibitem{GHJonker1966}
G. H. Jonker, 
J. Appl. Phys. \textbf{37}, 1424 (1966).

\bibitem{JBGoodenough2000}
J. B. Goodenough and R.I. Dass, 
Int. J. Inorg. Mater. \textbf{2}, 3 (2000).

\bibitem{BNRao2016}
B. N. Rao, L. Olivi, V. Sathe and R. Ranjan, 
Phys. Rev. B \textbf{93}, 024106 (2016).

\bibitem{CMeneghini2009}
C. Meneghini, Sugata Ray, F. Liscio, F. Bardelli, S. Mobilio, and D. D. Sarma, 
Phys. Rev. Lett \textbf{103}, 046403 (2009).

\bibitem{PSanyal2017}
P. Sanyal, 
Phys. Rev. B \textbf{96}, 214407 (2017).

\bibitem{FFan2016}
F. Fan, Z. Li, Z. Zhao, K. Yang and H. Wu, 
Phys. Rev. B \textbf{94}, 214401 (2016).

\bibitem{MPSingh2009}
M. P. Singh, K. D. Truong, S. Jand and P. Fournier, 
Appl. Phys. Lett. \textbf{98}, 162506 (2011).

\bibitem{DGivordl1992}
D. Givord, M.E Rossignol and D.W. Taylor, 
J. Phys. IV France \textbf{02} C3-95-C3-104 (1992).

\bibitem{HKronmiilier1988}
H. Kronmiilier, K.-D. Durst, M. Sagawa, 
J. Magn. Magn. Mater. \textbf{74} 291-302 (1988).

\bibitem{DCKakarla2014}
D. C. Kakarla, K. M. Jyothinagaram, A. K. Das and V. Adyam, 
J. Am. Ceram. Soc. \textbf{97}, 2858-2866 (2014).

\bibitem{RYadava2015}
R. Yadava and S. Elizabeth, 
J. Appl. Phys. \textbf{117}, 053902 (2015).

\bibitem{MSuzuki2009}
M. Suzuki, S. I. Fullem, I. S. Suzuki, L. Wang and C.-J. Zhong, 
Phys. Rev. B \textbf{79}, 024418 (2009).

\bibitem{KHFischer1987}
K. H. Fischer and Z. Zippelius, 
Phys. Rev. B \textbf{35}, 7171 (1987).

\bibitem{deAlmeida1978}
J. R. L. de Almeida and D. J. Thouless, 
J. Phys. A: Math. Gen. \textbf{11} 983 (1978).

\bibitem{MGabay1981}
M. Gabay and G. Toulouse, 
Phys. Rev. Lett. \textbf{47}, 5 201-204 (1981).

\bibitem{SSahoo2003}
S. Sahoo, O. Petracic, W. Kleemann, P. Nordblad, S. Cardoso, and P. P. Freitas, 
Phys. Rev. B \textbf{67}, 214422 (2003).

\bibitem{LSLakshmi2011}
L. S. Lakshmi and A. K. Nigam, 
J. Phys.: Condens. Matter \textbf{23} 086006 (2011) and references therein.

\bibitem{YSun2003}
Y. Sun, M. B. Salamon, K. Garnier, and R. S. Averback, 
Phys. Rev. Lett. \textbf{91}, 167206 (2003).

\bibitem{MSasaki2005}
M. Sasaki, P. E. J$\ddot{o}$nsson, H. Takayama and H. Mamiya, 
Phys. Rev. B \textbf{71}, 104405 (2005).

\bibitem{MTripathi2019}
M. Tripathi, T. Chatterji, H. E. Fischer, R. Raghunathan, S. Majumder, R. J. Choudhary and D. M. Phase, 
Phys. Rev. B \textbf{99}, 014422 (2019).

\bibitem{JSBenitez2011}
J S\'{a}nchez-Ben\'{i}tez, M. J. Mart\'{i}nez-Lope, J. A. Alonso and J L Garc\'{i}a-Mu\~{n}oz, 
J. Phys.: Condens. Matter \textbf{23}, 226001 (2011).

\bibitem{SPal2019}
S. Pal, S. Jana, S. Govinda, B. Pal, S. Mukherjee, S. Keshavarz, D. Thonig, Y. Kvashnin, M. Pereiro, R. Mathieu, P. Nordblad, J. W. Freeland, O. Eriksson, O. Karis, and D. D. Sarma, 
Phys. Rev. B \textbf{100}, 045122 (2019).

\bibitem{KJonason1996}
K. Jonason, J. Mattsson, and P. Nordblad, 
Phys. Rev. Lett. \textbf{77}, 2562 (1996).

\end{thebibliography}

\end{document}


\title{Supplementary Informations: \\ 
Mapping the magnetic state as a function of anti-site disorder \\ in Sm$ _{2} $NiMnO$ _{6} $ double perovskite thin films}

\author{Supriyo Majumder$^{a}$}
\author{Malvika Tripathi$^{a}$}
\author{R. Raghunathan$^{a}$}
\author{P. Rajput$^{b}$}
\author{S. N. Jha$^{b}$}
\author{D. O. de Souza$^{c}$}
\author{L. Olivi$^{c}$}
\author{S. Chowdhury$^{a}$}
\author{R. J. Choudhary$^{a}$}\email{ram@csr.res.in}
\author{and D. M. Phase$^{a}$}
\renewcommand{\andname}{\ignorespaces}
\affiliation{$^{a}$UGC DAE Consortium for Scientific Research, Indore 452001, India}
\affiliation{$^{b}$Beamline Development and Application Section, Bhabha Atomic Research Centre, Mumbai 400085, India}
\affiliation{$^{c}$Elettra Sicrotrone Trieste S.C.p.A., SS 14-km 163.5, 34149 Basovizza, Italy}

\maketitle

\begin{table*}[h!]\renewcommand{\thetable}{S\arabic{table}}
	\centering
\begin{tabular}{@{}|c|c|c|c|c|c|c|@{}}
	\hline
  Atoms & Site & \multicolumn{3}{c|}{Fractional coordinates} & Lattice parameters \\ 
   	 		\hline	
 	 &    		   & 	x 		& 	y 		& 	z 		& \\  			
  Sm & \textit{4e} & -0.0119(4) & 0.0563(1) & 0.250 & a = 5.35570(9) \\
   
  Ni & \textit{2c} & 0.500 & 0.000 & 0.500 & b = 5.53894(9) \\
  
  Mn & \textit{2d} & 0.500 & 0.000 & 0.000 & c = 7.61695(12) \\
   
  O1 & \textit{4e} & 0.0954(20) & 0.4795(18) & 0.2604(61) & $\alpha$=90.0 \\ 
  
  O2 & \textit{4e} & 0.6986(35) & 0.2879(36) & 0.0480(22) & $\beta$=90.032(2) \\  
  
  O3 & \textit{4e} & 0.7085(36) & 0.3037(34) & 0.4567(23) & $\gamma$=90.0 \\
  
		\hline \hline
	\end{tabular}
\caption{Crystallographic information obtained from Rietveld analysis of powder X-ray diffraction data recorded at T=300 K for polycrystalline SNMO sample.}
	\label{cif}
\end{table*} 

\begin{table*}[]\renewcommand{\thetable}{S\arabic{table}}
\centering
\begin{tabular}{@{}|c|c|c|c|c|c|c|c|c|c|@{}}
\hline
\multirow{2}{*}{Sample} & \multicolumn{2}{c|}{Ni-O} & \multicolumn{2}{c|}{Ni-Sm} & \multicolumn{2}{c|}{Ni-Mn} & \multicolumn{2}{c|}{Ni-Ni} & \multirow{2}{*}{q} \\ \cmidrule(lr){2-9} 
 & R (\AA) & $\sigma^{2}$ (10$^{-4}$\AA$^{2})$ & R (\AA) & $\sigma^{2}$ (10$^{-4}$\AA$^{2})$ & R (\AA) & $\sigma^{2}$ (10$^{-4}$\AA$^{2})$ & R (\AA) & $\sigma^{2}$ (10$^{-4}$\AA$^{2})$ &  \\ \hline
S\_HO & 2.055(3) & 67(5) & 3.087(6) & 129(9) & 3.962(8) & 107(8) & 3.953(8) & 107(8) & 0.95(1) \\

S\_I & 2.059(3) & 101(5) & 3.079(5) & 131(7) & 3.965(7) & 112(9) & 3.917(8) & 112(9) & 0.81(2) \\

S\_HD & 2.058(2) & 108(5) & 3.079(4) & 130(6) & 3.966(9) & 117(9) & 3.871(9) & 117(9) & 0.53(2) \\ 
\hline
\hline
\end{tabular}%
\caption{Results of Ni \textit{K}-edge extended X-ray absorption fine structure analysis, right most column reports estimated probability of getting ordered bond configurations present in the films. Number in the parentheses represent the error bars on the last digit.}
\label{EXAFS table}
\end{table*}

\begin{table*}[]\renewcommand{\thetable}{S\arabic{table}}
\centering
\begin{tabular}{|c|c|c|c|c|c|}
\hline
System     & Empty sample holder                & Bare STO substrate                          & S$ \_ $HD & S$ \_ $I & S$ \_ $HO \\ \hline
Moment (emu) & -7.55$ \times $10$ ^{-8}$ & -3.29$ \times $10$ ^{-7}$ & +4.53$ \times $10$ ^{-6}$ & +1.73$ \times $10$ ^{-5}$ & +5.61$ \times $10$ ^{-5}$ \\ 
		\hline \hline
\end{tabular}%
\caption{Observed moment values at T=10 K, in presence of $ \mu_{0} $H=100 Oe magnetic field.}
\label{tabresrawmoment10K}
\end{table*}

Table \ref{cif} lists the structural parameters determined from Rietveld analysis of X-ray diffraction data measured at T=300 K for polycrystalline SNMO sample.

\begin{figure*}[h!]\renewcommand{\thefigure}{S\arabic{figure}}
\centering
\includegraphics[angle=0,width=0.3\textwidth]{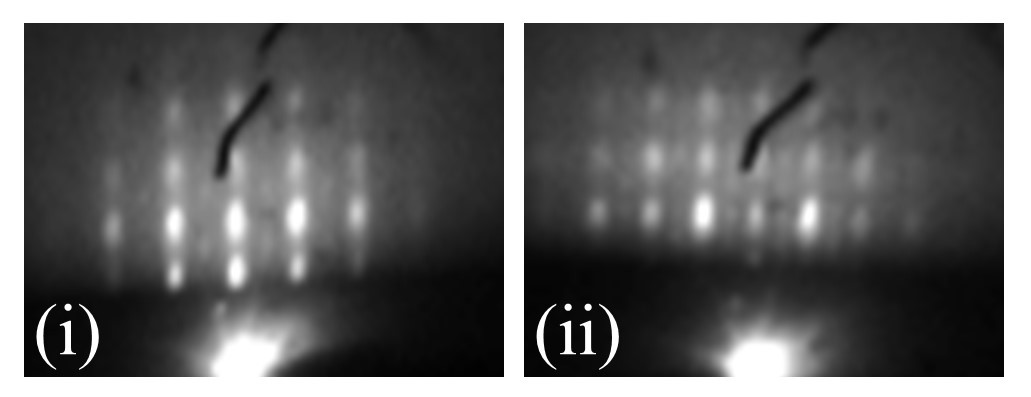}
\caption{Representative RHEED patterns for S$ \_ $I thin film surface measured at different angle of rotations.}\label{resrheedsiangle}
\end{figure*}

Representative RHEED patterns for SNMO S$ \_ $I thin film surface measured at different angles are shown in Figs. \ref{resrheedsiangle}.

\begin{figure*}[h!]\renewcommand{\thefigure}{S\arabic{figure}}
\centering
\includegraphics[angle=0,width=0.8\textwidth]{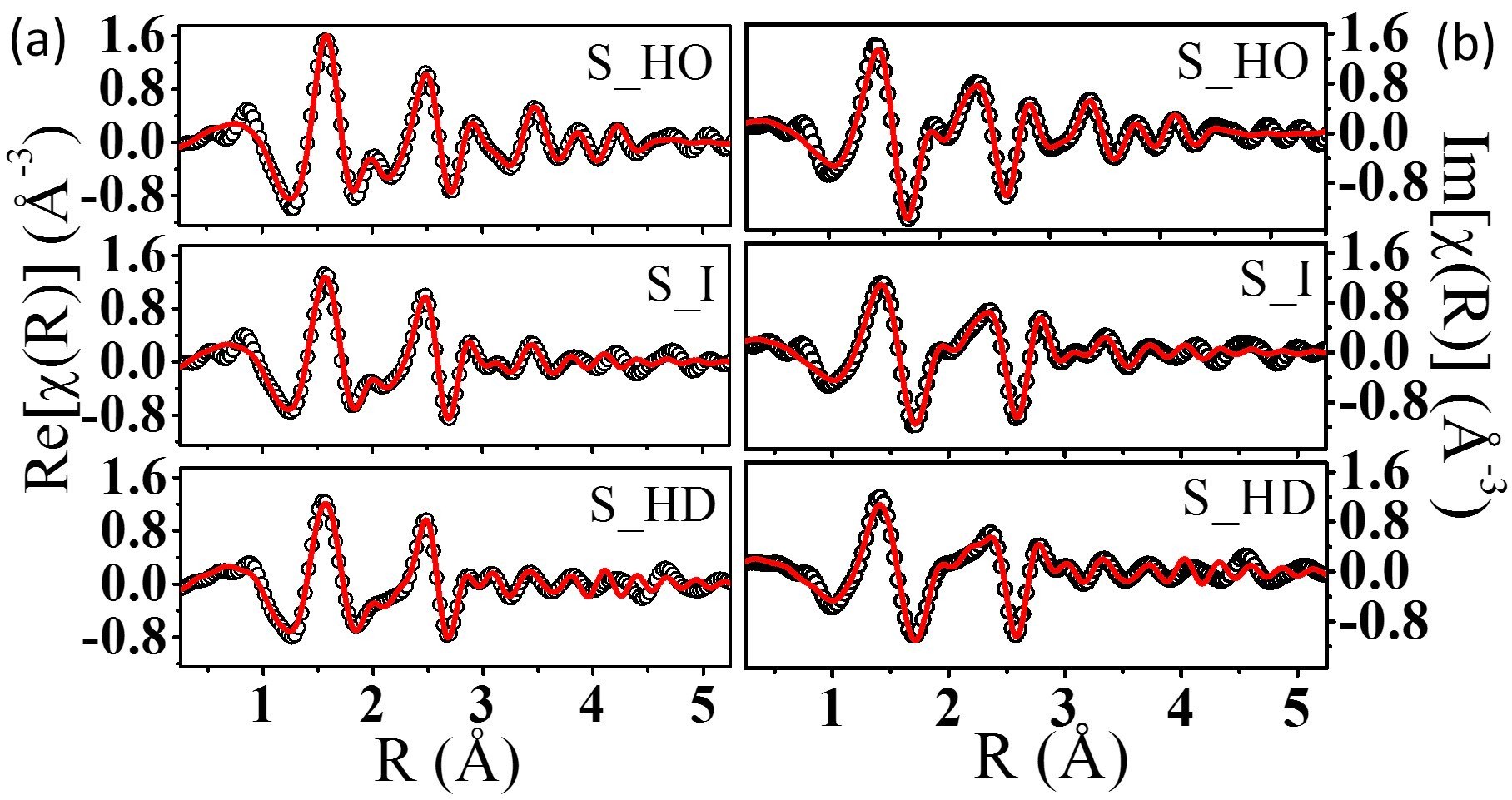}
\caption{(a): Real and (b): Imaginary parts of Fourier transformed $k$-space Ni \textit{K}-edge extended X-ray absorption fine structure spectra for SNMO samples presenting observed signals (hollow/black circles) and corresponding best fits (solid/red lines).}\label{SMReIMFT}
\end{figure*}

Figures \ref{SMReIMFT}(a, b) display observed signals (hollow/black circles) and corresponding best fits (solid/red lines) in real and imaginary part of Fourier transformed $k$-space Ni \textit{K}-edge extended X-ray absorption fine structure (EXAFS) spectra for SNMO samples. Table \ref{EXAFS table} lists the refined values of structural parameters obtained from EXAFS modeling for the SNMO films. 

\begin{figure*}[h!]\renewcommand{\thefigure}{S\arabic{figure}}
\centering
\includegraphics[angle=0,width=1.0\textwidth]{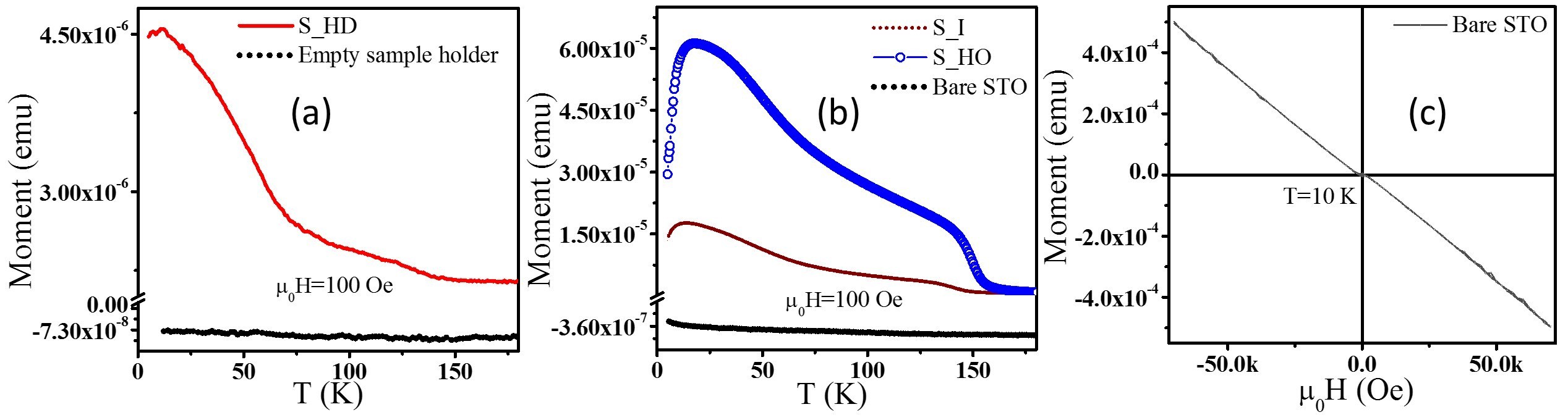}
\caption{Field cooled warming moment as a function of temperature in presence of 100 Oe applied magnetic field, for (a): empty sample holder, S$ \_ $HD film and for (b): bare STO substrate, S$ \_ $I, S$ \_ $HO films. (c): Isothermal magnetization measured as a function of magnetic filed at T=10 K for bare STO substrate. Film and substrate magnetization are measured along in-plane geometry.}\label{resrawmoment}
\end{figure*}

The magnetization behaviors for empty sample holder, bare STO substrate and SNMO thin films are compared in Fig.\ref{resrawmoment}. Film and substrate magnetization are recorded along in-plane geometry. Observed moment values at T=10 K, measured under $ \mu_{0} $H=100 Oe of applied magnetic field are listed in Table. \ref{tabresrawmoment10K}. 

\begin{figure*}[h!]\renewcommand{\thefigure}{S\arabic{figure}}
\centering
\includegraphics[angle=0,width=0.60\textwidth]{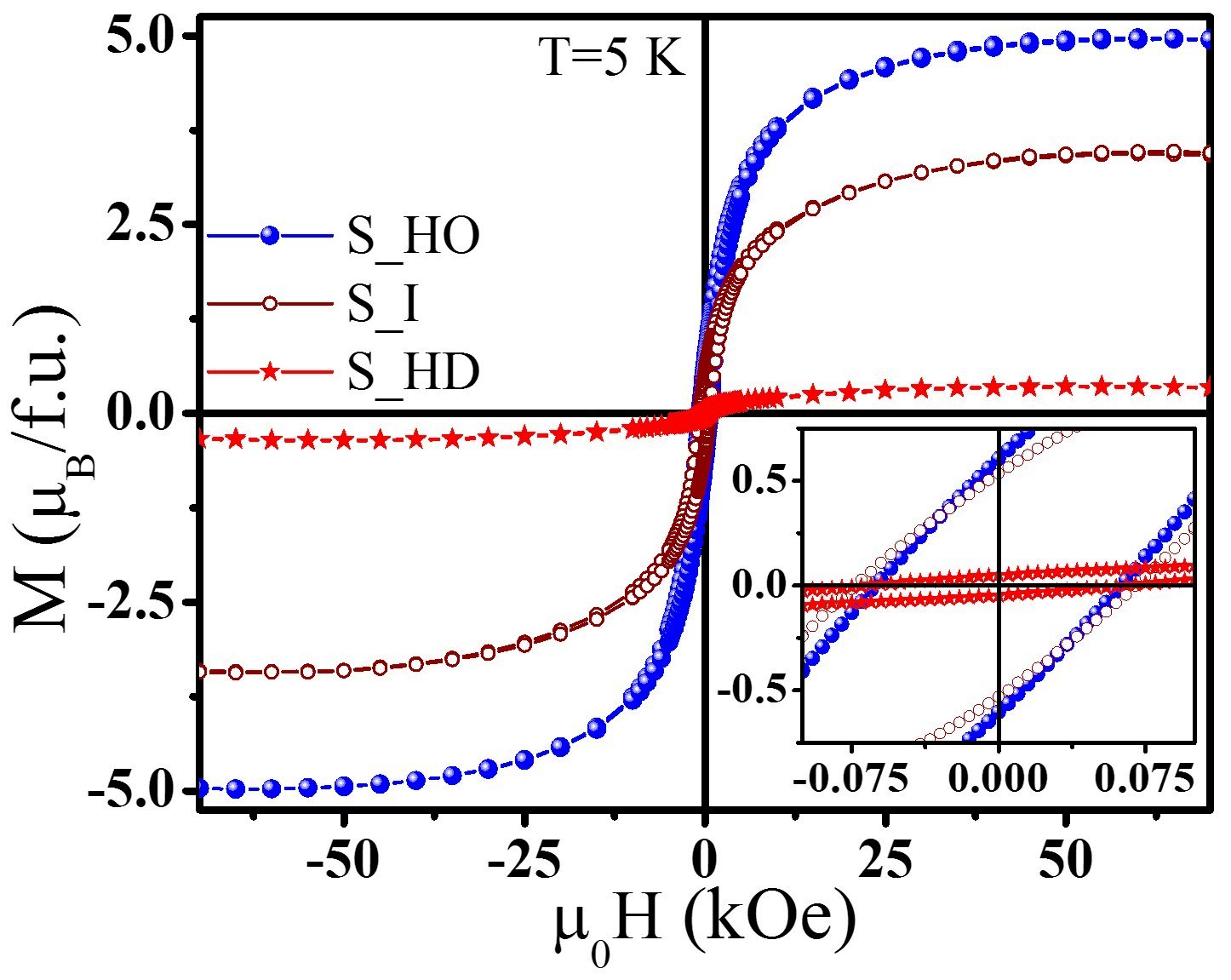}
\caption{In-plane magnetization measured as a function of applied field at T=5 K for SNMO films having different anti-site disorder densities. Inset show enlarged view of low field region.}\label{SMMH}
\end{figure*}

Isothermal magnetization measured as a function of applied magnetic field M(H) at T=5 K temperature for SNMO films having different ASD concentrations are presented in Fig. \ref{SMMH}. Here M(H) behaviors are recorded along in-plane geometry. 

\begin{figure*}[h!]\renewcommand{\thefigure}{S\arabic{figure}}
\centering
\includegraphics[angle=0,width=0.85\textwidth]{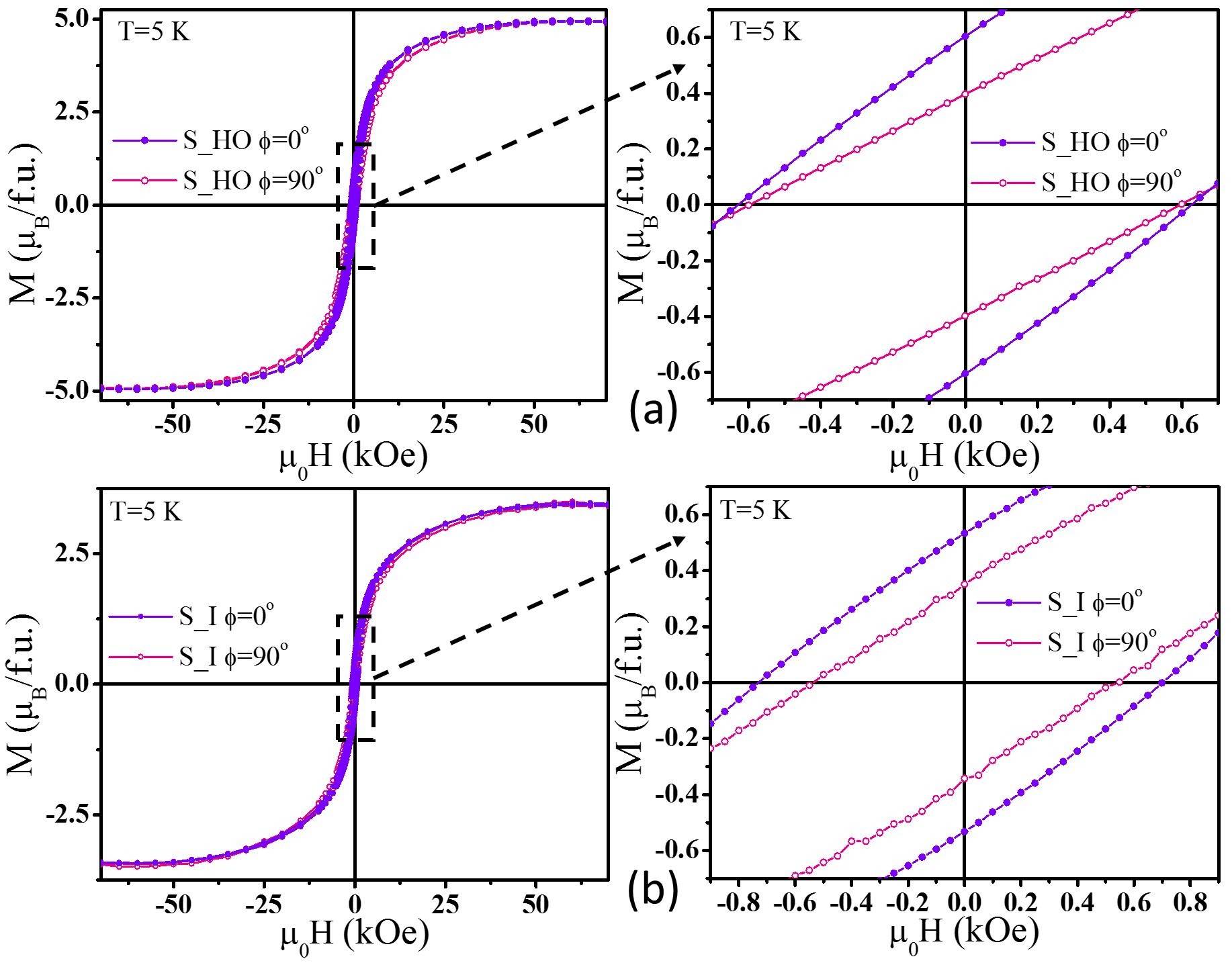}
\includegraphics[angle=0,width=0.85\textwidth]{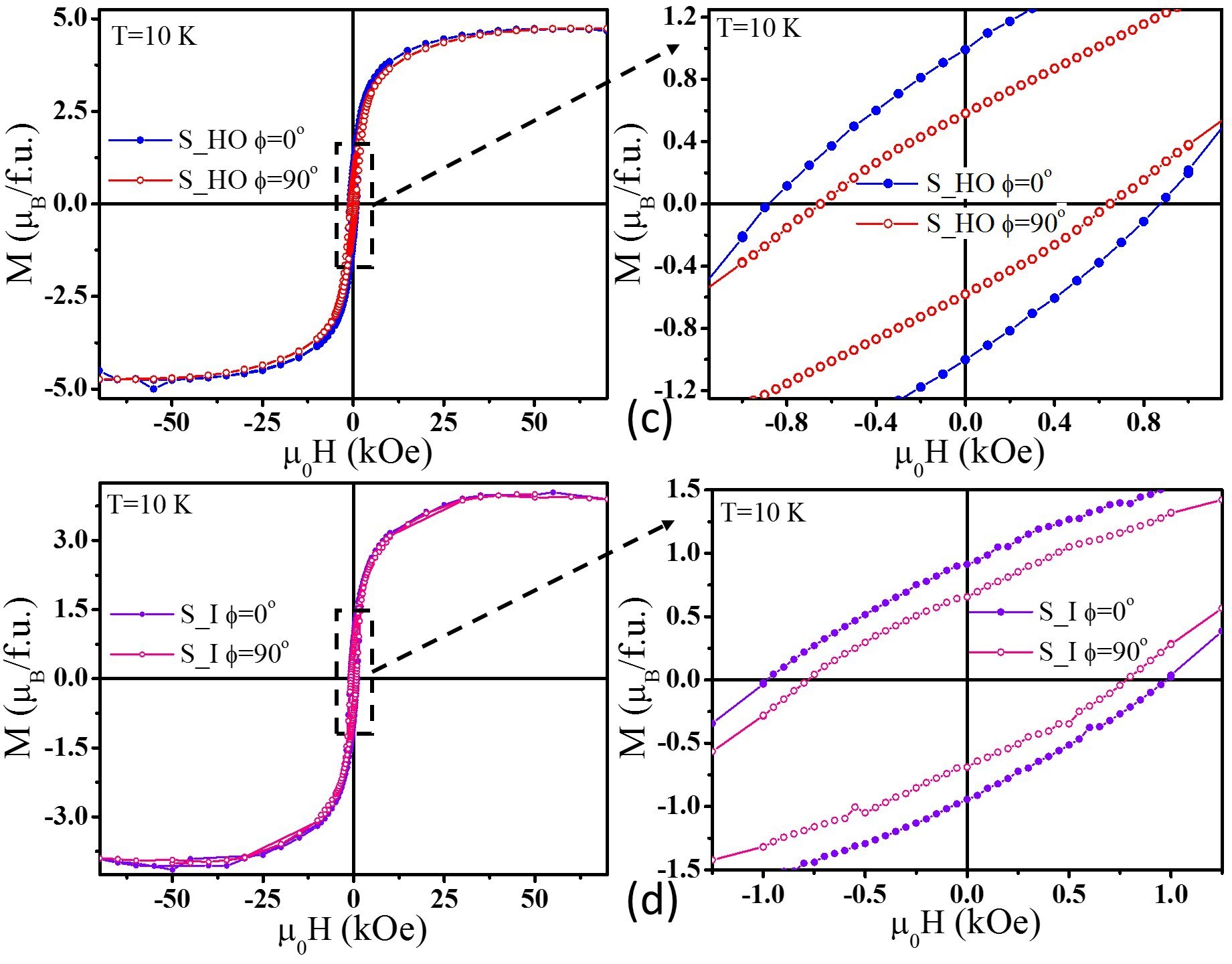}
\caption{In-plane and out-of-plane isothermal magnetizations recorded as a function of applied field (a): at T=5 K for S$ \_ $HO film, (b): at T=5 K for S$ \_ $I film, (c): at T=10 K for S$ \_ $HO film and (d): at T=10 K for S$ \_ $I film. The right panels show enlarged view in low field region of corresponding left panel plots.}\label{smmh5and10kipop}
\end{figure*}

\begin{figure*}[h!]\renewcommand{\thefigure}{S\arabic{figure}}
\centering
\includegraphics[angle=0,width=0.85\textwidth]{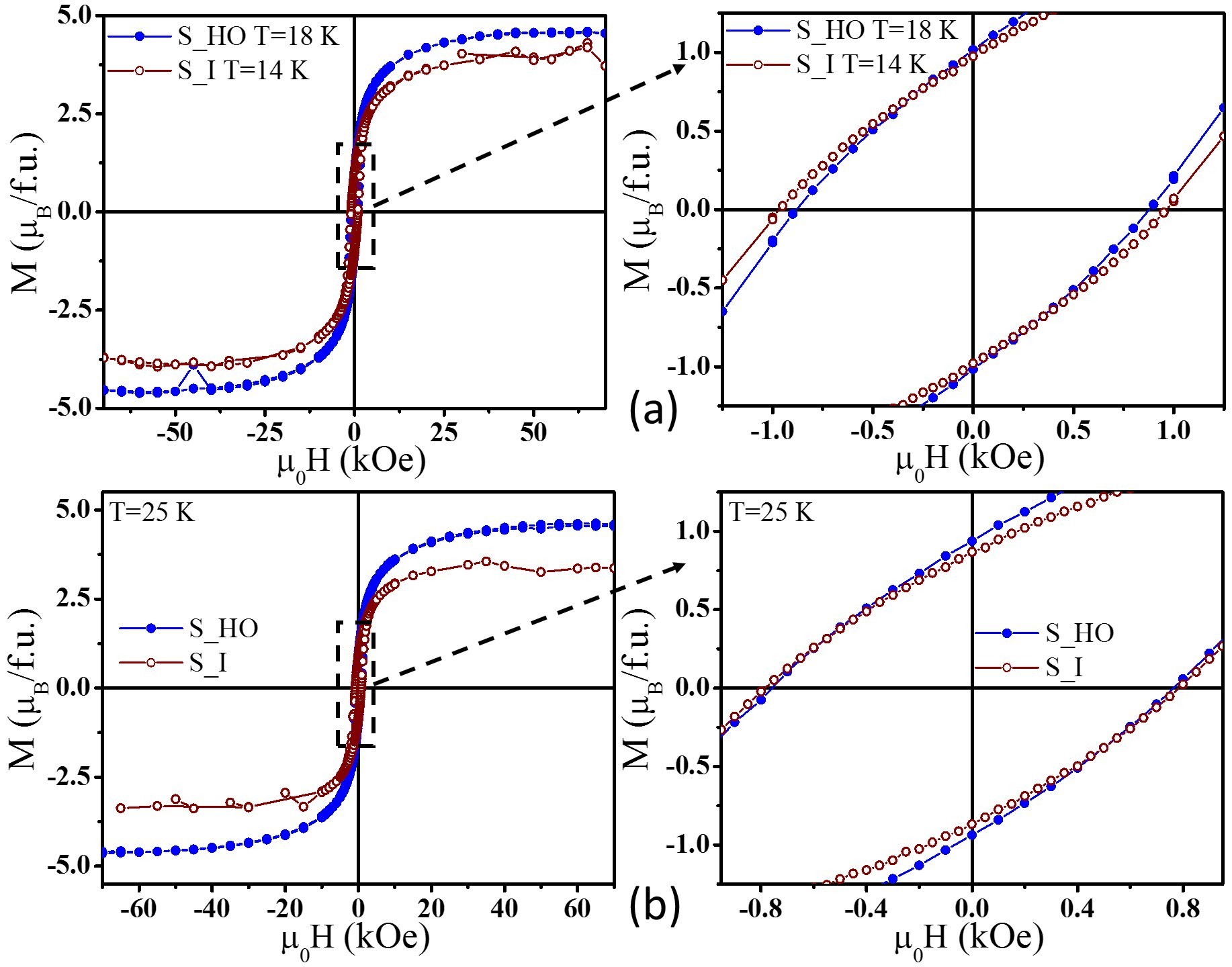}
\includegraphics[angle=0,width=0.85\textwidth]{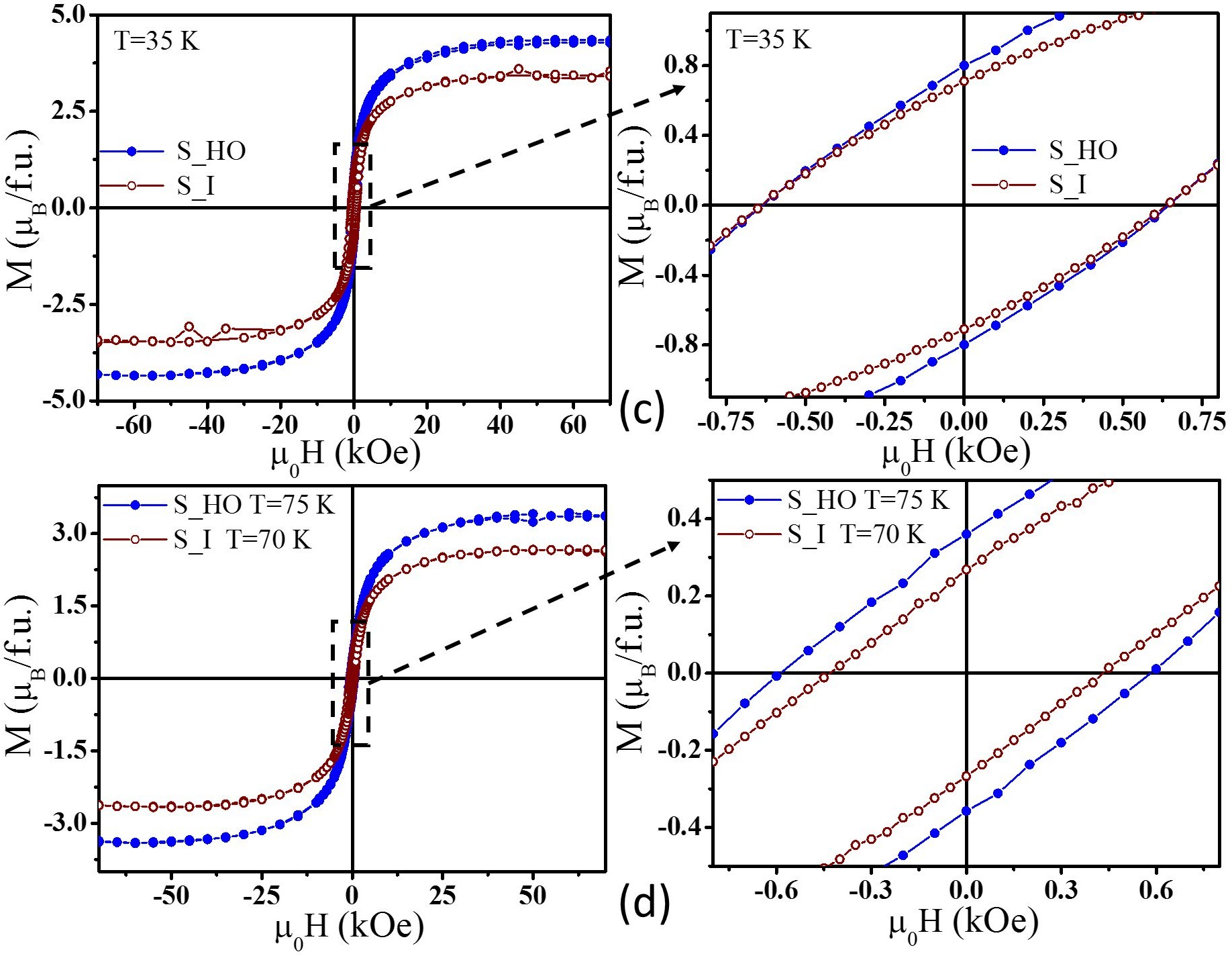}
\caption{In-plane isothermal magnetizations recorded as a function of applied field (a): at T=18 K for S$ \_ $HO film, at T=14 K for S$ \_ $I film, (b): at T=25 K for S$ \_ $HO film, at T=25 K for S$ \_ $I film, (c): at T=35 K for S$ \_ $HO film, at T=35 K for S$ \_ $I film and (d): at T=75 K for S$ \_ $HO film, at T=70 K for S$ \_ $I film. The right panels show enlarged view in low field region of corresponding left panel plots.}\label{smmh18and25and35and75k}
\end{figure*}

\begin{figure*}[h!]\renewcommand{\thefigure}{S\arabic{figure}}
\centering
\includegraphics[angle=0,width=0.85\textwidth]{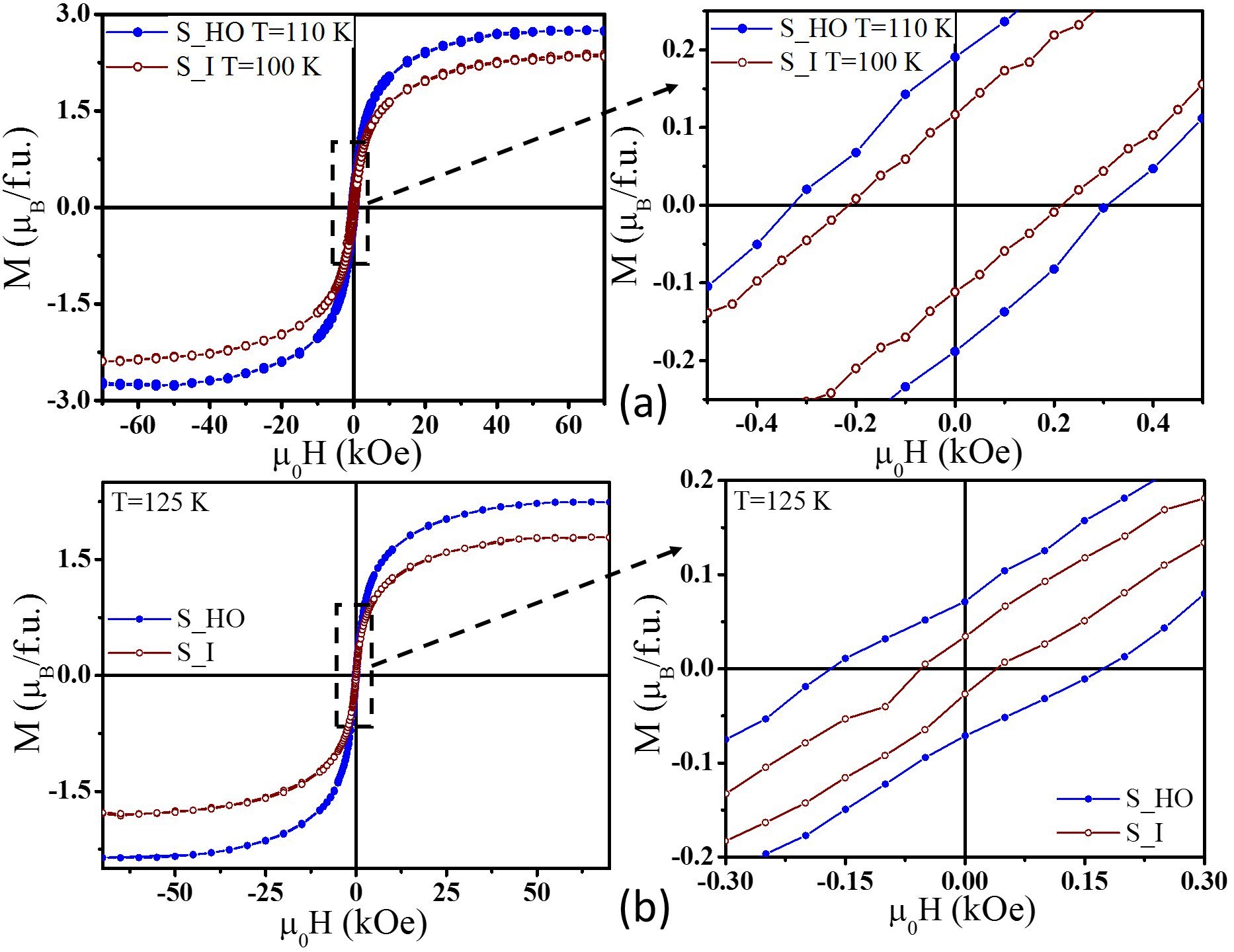}
\caption{In-plane isothermal magnetizations recorded as a function of applied field (a): at T=110 K for S$ \_ $HO film, at T=100 K for S$ \_ $I film and (b): at T=125 K for S$ \_ $HO film, at T=125 K for S$ \_ $I film. The right panels show enlarged view in low field region of corresponding left panel plots.}\label{smmh110and125k}
\end{figure*}

M(H) curves recorded in different orientations with respect to applied field directions and at different temperature values are displayed in Figs. \ref{smmh5and10kipop}, \ref{smmh18and25and35and75k}, \ref{smmh110and125k}.

\begin{figure*}[h!]\renewcommand{\thefigure}{S\arabic{figure}}
\centering
\includegraphics[angle=0,width=0.5\textwidth]{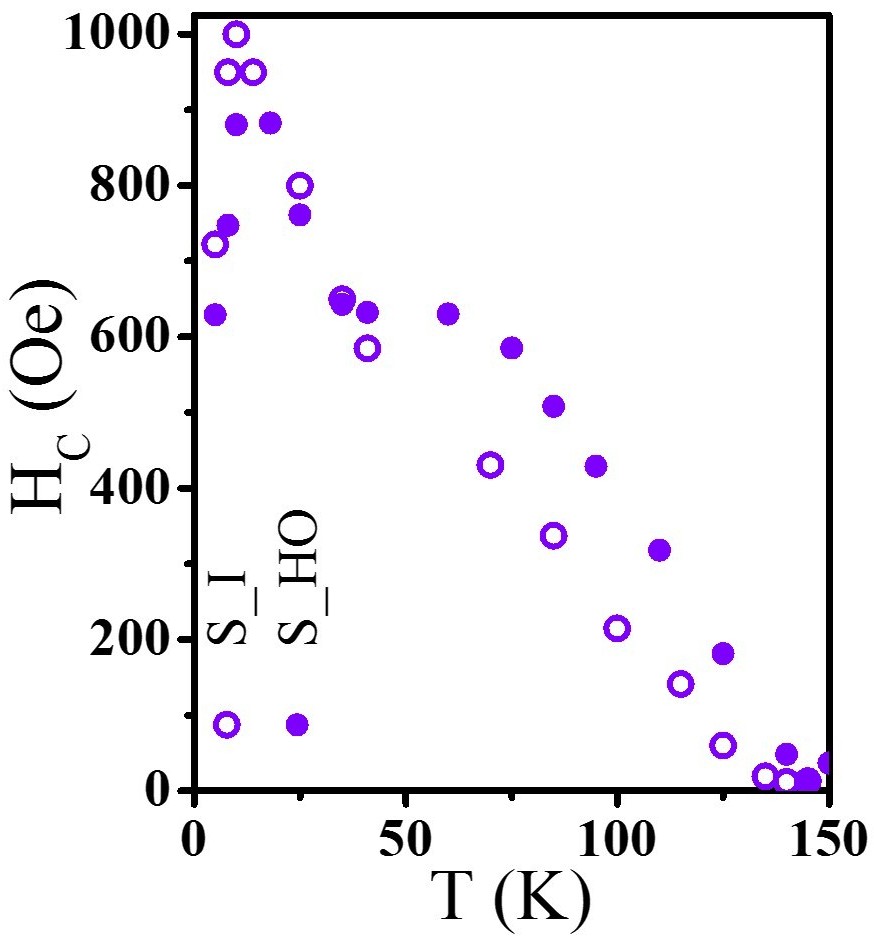}
\caption{Thermal evolution of coercivity measured along in-plane geometic configuration for SNMO thin films.}\label{hct}
\end{figure*}

The temperature dependency of coercivity H$ _{C} $(T) measured for SNMO films are presented in Fig. \ref{hct}. 

\begin{figure*}[h!]\renewcommand{\thefigure}{S\arabic{figure}}
\centering
\includegraphics[angle=0,width=0.95\textwidth]{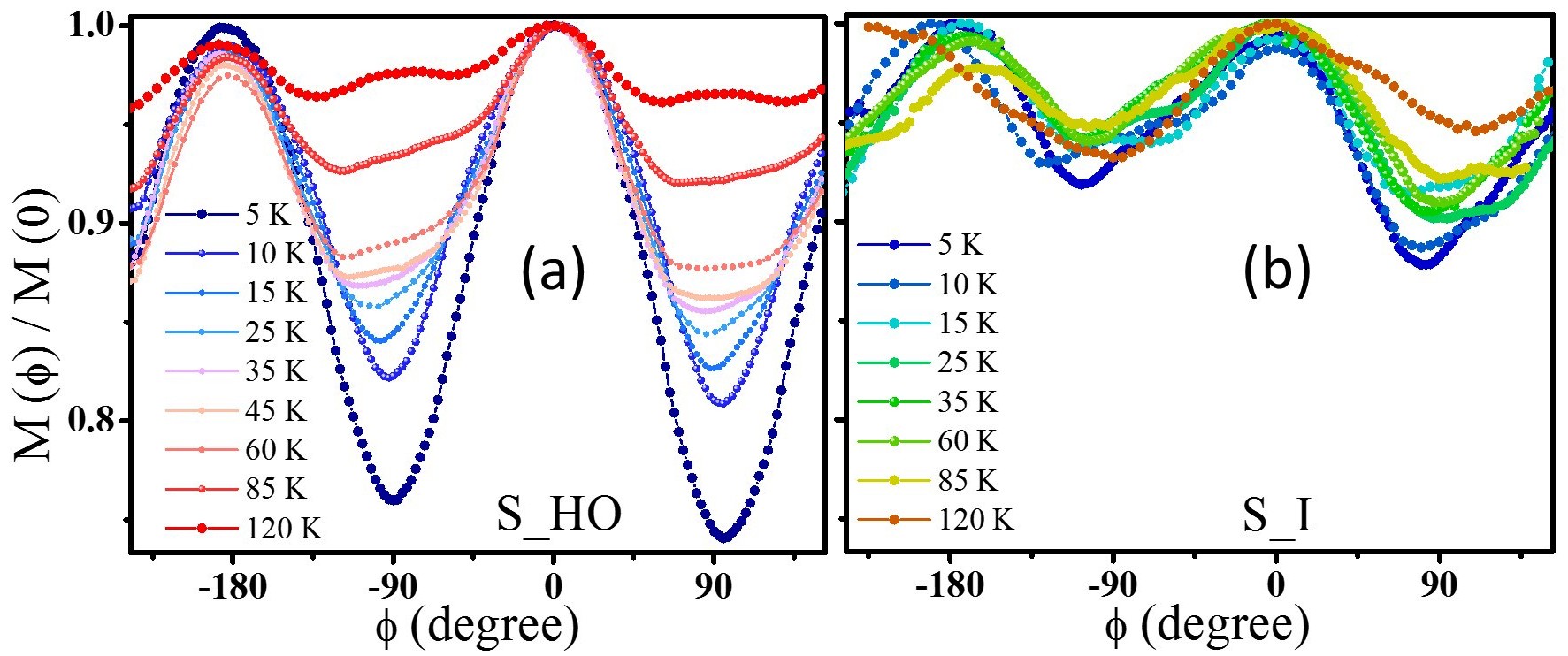}
\caption{The angular dependency of magnetization at different temperatures in presence of $ \mu_{0} $H=2500 Oe for (a): S$ \_ $HO and (b): S$ \_ $I films.}\label{mth}
\end{figure*}

Figures \ref{mth}(a, b) illustrate the angular variation of magnetization recorded at different temperatures in presence of $ \mu_{0} $H=2500 Oe for SNMO films.

\begin{figure*}[h!]\renewcommand{\thefigure}{S\arabic{figure}}
\centering
\includegraphics[angle=0,width=0.95\textwidth]{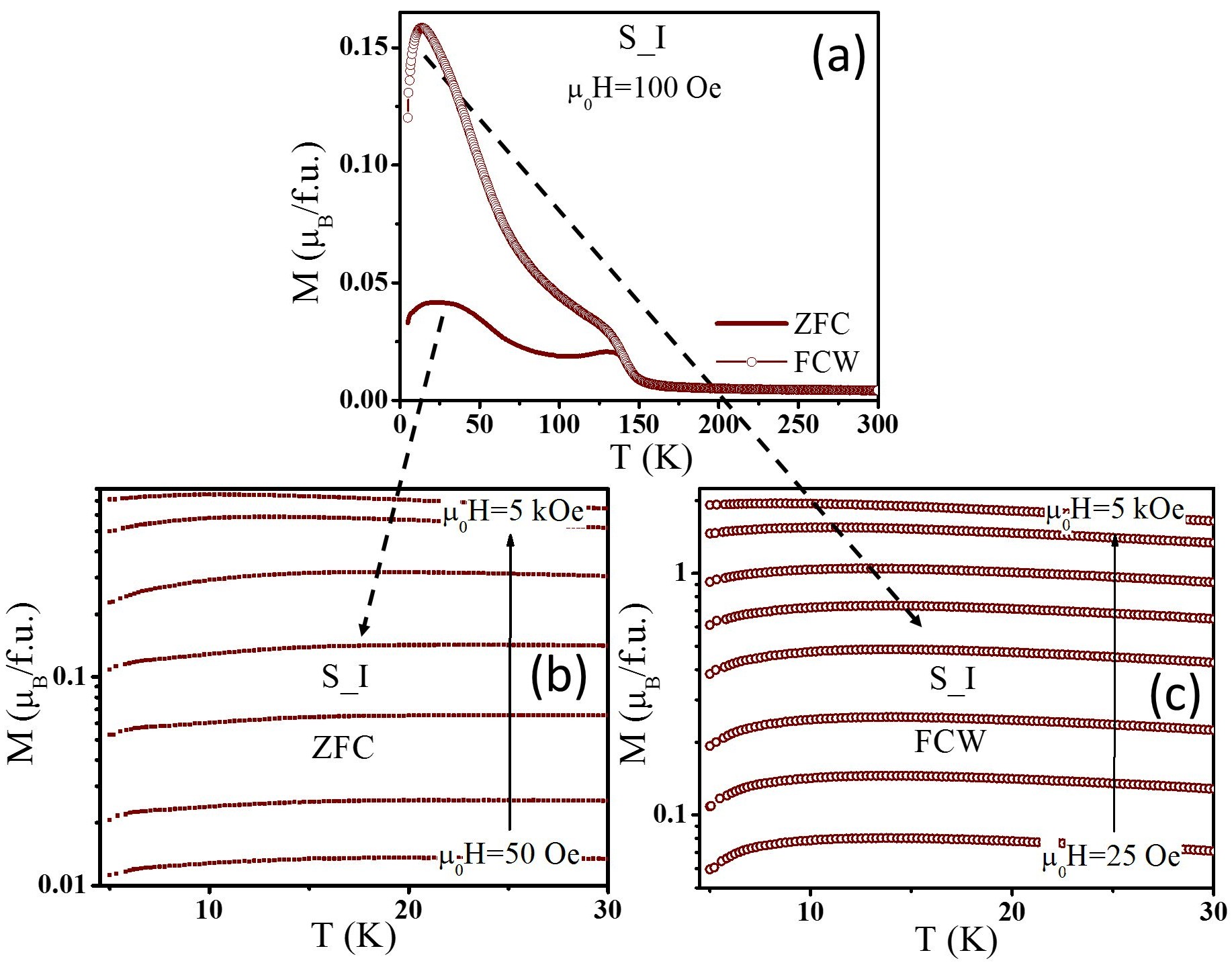}
\includegraphics[angle=0,width=0.55\textwidth]{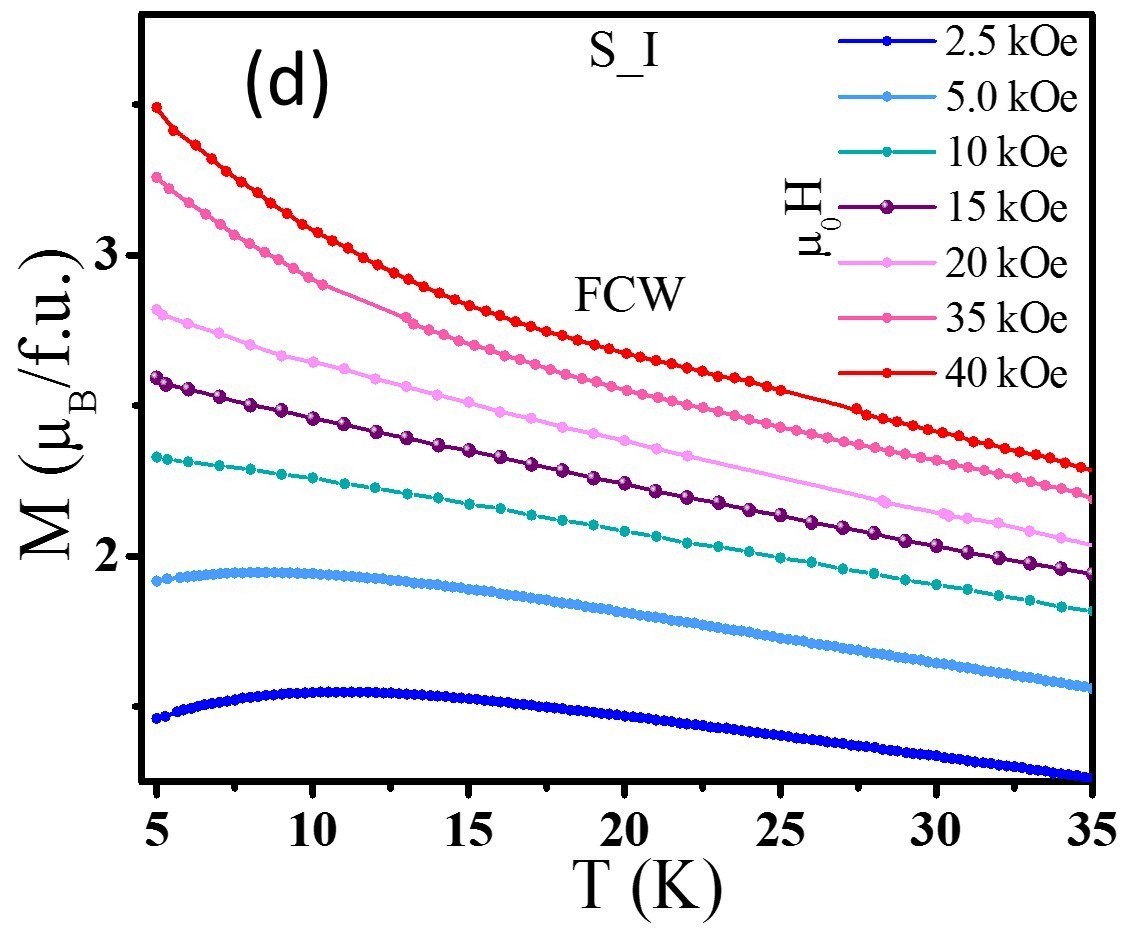}
\caption{(a): Magnetization measured as a function of temperature M(T) in presence of external magnetic field 100 Oe, following ZFC and FCW protocols, for S$\_$I sample. M(T) curves recorded under different measuring fields, in (b): ZFC and (c, d): FCW modes. All of these magnetization are measured along in-plane geometry.}\label{S_I MT}
\end{figure*}

Temperature dependent magnetization for S$\_$I thin film measured under 100 Oe applied magnetic field in zero field cooled (ZFC) warming and field cooled warming (FCW) paths are displayed in Fig.\ref{S_I MT}(a). Below downturn transition temperature T$ _{d} $, both ZFC and FCW magnetization are found to decrease. To explore the effect of measuring field on downturn transition, M(T) curves are recorded for different applied field values in both ZFC and FCW modes, as presented in Fig.\ref{S_I MT}(b, c and d). All of these magnetization behaviors are recorded along in-plane geometry. 

